\documentclass[12pt,eqno,epsf]{article}
\usepackage{amsmath,amssymb,graphicx,slashbox}
\numberwithin{equation}{section}

\def\mydate{May 18, 2007}
\def\ignore#1{{}}

\newcounter{sxn}

\newcounter{axn}

\date{}

\newdimen\mybaselineskip
\renewcommand{\thefootnote}{\arabic{footnote}}

\newcommand{\beeq}{\begin{equation}}
\newcommand{\eneq}{\end{equation}}
\newcommand{\beqn}{\begin{eqnarray}}
\newcommand{\eeqn}{\end{eqnarray}}

\newcommand{\alp}{\alpha}
\newcommand{\bt}{\beta}
\newcommand{\gm}{\gamma}
\newcommand{\Gm}{\Gamma}
\newcommand{\dlt}{\delta}

\newcommand{\ep}{\epsilon}
\newcommand{\vep}{\varepsilon}
\newcommand{\tht}{\theta}

\newcommand{\lmd}{\lambda}
\newcommand{\Lmd}{\Lambda}
\newcommand{\sgm}{\sigma}

\newcommand{\vph}{\varphi}
\newcommand{\omg}{\omega}
\newcommand{\Omg}{\Omega}

\newcommand{\be}{\begin{equation}}
\newcommand{\ee}{\end{equation}}
\newcommand{\bea}{\begin{eqnarray}}
\newcommand{\eea}{\end{eqnarray}}
\newcommand{\eql}{\!\!\!&=\!\!\!&}

\newcommand{\sqa}{\!\!\!&\simeq\!\!\!&}
\newcommand{\defa}{\!\!\!&\equiv\!\!\!&}

\newcommand{\toa}{\!\!\!&\to\!\!\!&}
\newcommand{\mtrx}[4]{\begin{pmatrix}#1&#2\\#3&#4\end{pmatrix}}
\newcommand{\vct}[2]{\begin{pmatrix}#1\\#2\end{pmatrix}}

\newcommand{\simlt}{\stackrel{<}{{}_\sim}}

\newcommand{\tl}[1]{\tilde{#1}}
\newcommand{\bdm}[1]{{\mbox{\boldmath $#1$}}}
\newcommand{\tr}{{\rm tr}}

\newcommand{\diag}{{\rm diag}}
\newcommand{\der}{\partial}
\newcommand{\dr}{\!\!d}

\newcommand{\ie}{{\it i.e.}}

\newcommand{\vev}[1]{\langle #1 \rangle}

\newcommand{\brkt}[1]{\left( #1 \right)}
\newcommand{\brc}[1]{\left\{ #1 \right\}}
\newcommand{\sbk}[1]{\left[ #1 \right]}
\newcommand{\abs}[1]{\left| #1 \right|}

\newcommand{\cA}{{\cal A}}
\newcommand{\cB}{{\cal B}}
\newcommand{\cC}{{\cal C}}
\newcommand{\cD}{{\cal D}}
\newcommand{\cF}{{\cal F}}

\newcommand{\cL}{{\cal L}}

\newcommand{\cO}{{\cal O}}
\newcommand{\cP}{{\cal P}}
\newcommand{\cQ}{{\cal Q}}

\newcommand{\cW}{{\cal W}}
\newcommand{\cZ}{{\cal Z}} 
\newcommand{\bG}{{\mathbb G}}
\newcommand{\bH}{{\mathbb H}}

\newcommand{\zp}{z_\pi}
\newcommand{\thw}{\tht_W}
\newcommand{\thH}{\tht_{\rm H}}
\newcommand{\cw}{\bar{c}_{\rm H}}
\newcommand{\sw}{\bar{s}_{\rm H}}
\newcommand{\cth}{c_{\rm H}}
\newcommand{\sth}{s_{\rm H}}
\newcommand{\cph}{c_\phi}
\newcommand{\sph}{s_\phi}

\newcommand{\ubl}{U(1)_{\rm B-L}}
\newcommand{\suL}{SU(2)_{\rm L}}
\newcommand{\suR}{SU(2)_{\rm R}}
\newcommand{\uy}{U(1)_Y}
\newcommand{\uem}{U(1)_{\rm EM}}
\newcommand{\mKK}{m_{\rm KK}}
\newcommand{\aL}{a_{\rm L}}
\newcommand{\aR}{a_{\rm R}}

\newcommand{\chL}{\pm_{\rm L}}
\newcommand{\chR}{\pm_{\rm R}}

\ignore{

}
\newcommand{\ha}{\hat{a}}
\newcommand{\fH}{f_{\rm H}}
\newcommand{\bthH}{\bar{\theta}_{\rm H}}

\newcommand{\NP}[1]{{\it Nucl.~Phys.}~{\bf #1}}
\newcommand{\PL}[1]{{\it Phys.~Lett.}~{\bf #1}}

\newcommand{\MPL}[1]{{\it Mod.~Phys.~Lett.}~{\bf #1}}

\newcommand{\PR}[1]{{\it Phys.~Rev.}~{\bf #1}}
\newcommand{\PRL}[1]{{\it Phys.~Rev.~Lett.}~{\bf #1}}
\newcommand{\PTP}[1]{{\it Prog.~Theor.~Phys.}~{\bf #1}}

\newcommand{\JH}[1]{{\it JHEP}~{\bf #1}}

\begin{document}
\thispagestyle{empty}

\baselineskip=12pt

{\small \noindent \mydate    \hfill OU-HET 581/2007}


\baselineskip=35pt plus 1pt minus 1pt

\vskip 2.0cm

\begin{center}
{\Large \bf Effective theories of gauge-Higgs unification models}\\
{\Large \bf in warped spacetime}\\

\vspace{2.0cm}
\baselineskip=20pt plus 1pt minus 1pt

{\def\thefootnote{\fnsymbol{footnote}}
\bf 
Yutaka\ Sakamura\footnote[1]{sakamura@het.phys.sci.osaka-u.ac.jp}
}\\
\vspace{.3cm}
{\small \it Department of Physics, Osaka University,
Toyonaka, Osaka 560-0043, Japan}\\
\end{center}

\vskip 2.0cm
\baselineskip=20pt plus 1pt minus 1pt

\begin{abstract}
We derive four-dimensional (4D) effective theories of 
the gauge-Higgs unification models in the warped spacetime. 
The effective action can be expressed in a simple form 
by neglecting subleading corrections to the wave functions. 
We have shown in our previous works that some Higgs couplings 
to the other fields are suppressed by factors that depend on $\bthH$ 
from the values in the standard model. 
Here $\bthH$ is the Wilson line phase along the fifth dimension, 
which characterizes the electroweak symmetry breaking. 
The effective action derived here explicitly shows 
a nonlinear structure of the Higgs sector, 
which clarifies the origins of those suppression factors. 
\ignore{
The derivation is much simpler than 
the conventional Kaluza-Klein (KK) expansion analysis 
and the results are completely consistent with 
those obtained by the KK analysis. 
The Higgs sector has a nonlinear structure, 
which leads to deviation of some couplings from 
the standard model values. 
The derived 4D actions have similar forms to those of 
4D models in which the Higgs bosons are provided as 
the pseudo Nambu-Goldstone bosons. 
This is consistent with the results obtained by 
the holographic approach.  }
\end{abstract}


\newpage
\section{Introduction}
In spite of many successes of the standard model, 
it is not believed to be the final theory because of 
some theoretical problems it has. 
One of the problems is an instability of the Higgs boson mass 
against radiative corrections. 
It has been argued that the Higgs mass suffers from 
quadratically divergent radiative corrections, 
which requires unnatural fine tuning of parameters 
in the theory unless it is protected by some symmetry. 
Supersymmetry is one of the most promising candidate 
of such symmetry. 
The supersymmetric models generally predict a light Higgs boson. 
In the minimal supersymmetric standard model, for example,  
the upper bound for the Higgs boson mass 
is about 130~GeV~\cite{OYY}, which is relatively close to 
the experimental lower bound~114~GeV~\cite{LEPgroup}. 
In recent years many alternative scenarios for the Higgs sector 
have been proposed, such as the little Higgs model~\cite{littleHiggs}, 
the Higgsless model~\cite{Higgsless}, and so on. 
Among them the gauge-Higgs unification (GHU) scenario 
predicts various interesting properties 
in the Higgs couplings to the gauge and the fermion fields. 

In the GHU scenario the Higgs field is unified with 
the gauge fields within the framework of higher dimensional 
gauge theory~\cite{Manton}-\cite{HS2}. 
The extra-dimensional components of 
the gauge potentials play a role of the Higgs scalar fields 
in four-dimensional (4D) effective theory. 
The electroweak symmetry can be broken dynamically 
by non-Abelian Aharonov-Bohm phases (Wilson line phases) 
when the extra-dimensional space is 
nonsimply-connected~\cite{Hosotani}-\cite{Haba}. 
This Hosotani mechanism also provides 
a finite mass to the 4D Higgs scalar by quantum dynamics. 
The Higgs mass is protected against the large radiative 
corrections by the higher dimensional gauge symmetry~\cite{finite_mass}. 

The simplest setup for the GHU scenario is 
five-dimensional (5D) gauge theory whose fifth dimension is 
compactified on an orbifold~$S^1/Z_2$~\cite{Hatanaka}, 
which naturally realizes chiral fermions in low energies. 
In the case that the 5D spacetime is flat, 
it has been shown that the Higgs mass is 
too small to satisfy the experimental lower bound 
unless the Wilson line phase~$\bthH$ is dynamically determined 
to a tiny value.\footnote{
This is a generic feature of the GHU models in the flat spacetime. 
See Ref.~\cite{HNT}, for example. } 
Besides, we have shown in Ref.~\cite{HS1} 
that trilinear couplings among 
the $W$ and the $Z$ bosons substantially deviate 
from the standard model values if $\bthH=\cO(1)$, 
which is inconsistent with the experiments. 
The warped Randall-Sundrum spacetime~\cite{RS} 
ameliorates these problems. 
The Higgs mass is enhanced by a factor~$k\pi R\simeq 35$ 
compared to the case of the flat spacetime~\cite{HM}
where $e^{k\pi R}$ is the warp factor that is used to explain 
a large hierarchy between the electroweak scale and 
the Planck scale. 
The couplings among the $W$ and the $Z$ bosons are 
in good agreement with those in the standard model 
even in the case that $\bthH=\cO(1)$~\cite{SH,HS1}. 
Therefore we consider the 5D GHU models in the warped spacetime 
whose fifth dimension is compactified on $S^1/Z_2$ in this paper. 

The conventional method to analyze models with an extra dimension 
is the so-called Kaluza-Klein (KK) expansion analysis. 
The procedure is as follows. 
Firstly we expand the 5D fields into infinite 
4D KK modes by means of some complete sets of functions of 
the extra-dimensional coordinate~$y$. 
The 4D modes become mass-eigenstates if we 
choose the complete sets properly. 
Such functions of $y$ are called the mode functions. 
The boundary conditions at both orbifold boundaries 
determine the mass spectrum and the mode functions. 
Substituting the mode-expanded expressions of the 5D fields 
into the 5D action and perform the $y$-integral, 
we obtain the 4D action. 
By means of this KK analysis we investigated 
the GHU models in the warped spacetime 
in our previous papers~\cite{HNSS,SH,HS1}. 
We have found that the Higgs couplings to the other fields 
deviate from their counterparts in the standard model. 
Namely the former are suppressed by factors that depend 
on $\bthH$. 
This indicates that the Higgs sector has a nonlinear 
structure in the low-energy effective theory. 

An alternative approach to analyze the GHU models is 
the so-called holographic approach proposed by Ref.~\cite{LPR}, 
which is inspired by the AdS/CFT correspondence~\cite{AdSCFT}.  
In this approach the field values on one boundary are treated as 
independent degrees of freedom from the bulk, and the 4D action 
is obtained by integrating out the latter. 
This approach is powerful for certain purposes, 
for example, the calculation of the Higgs potential or 
of the electroweak oblique parameters~\cite{PT}. 
Such quantities can be estimated 
without explicitly summing contributions 
from the infinite KK modes~\cite{warpGHU:so5,holograph,AC}. 
Recently the authors of Ref.~\cite{Panico} derived 
the 4D effective actions of the 5D gauge theories 
including the GHU models in the holographic approach. 
They showed how to incorporate the 4D scalars coming from 
the fifth components of the 5D gauge fields 
into the 4D action. 
This work clarifies the symmetry structure 
of the effective action. 
The effective action derived in the holographic approach 
is expressed in the momentum space and 
in terms of the boundary values 
of the 5D fields, which are not mass-eigenstates. 
Thus the evaluation of the cubic (quartic) couplings requires 
calculations of the three- (four-) point functions. 
One of the main purposes of this paper 
is to understand the suppression factors 
for the Higgs couplings directly from the nonlinear structure 
of the Higgs sector. 
The KK analysis is suitable for this purpose 
because the derived 4D action is expressed 
in terms of the mass-eigenstates 
and the coupling constants are directly read off from it. 
So we apply the latter approach to derive the effective action 
in this paper. 

\ignore{
Although the KK analysis is systematic and powerful to analyze 
the spectrum and the coupling constants, 
it involves somewhat complicated calculations. }
In the warped spacetime the mode functions are written in terms of 
the Bessel functions and the mass eigenvalues cannot be 
expressed in analytic forms. 
For the light modes below the KK mass scale~$\mKK$, however, 
they have simple approximate expressions that are 
analytic functions of $\bthH$~\cite{HNSS,SH,HS1}. 
Correction terms to them are suppressed 
by $\cO(\pi^2 m_f^2/\mKK^2)$ 
where $m_f$ are masses of the light modes. 
This indicates that we can obtain a simple form of 
the effective action by neglecting the subleading contributions 
of $\cO(\pi^2 m_f^2/\mKK^2)$. 
Besides, it is convenient to gauge away 
the background of the gauge fields~$A_y^{\rm bg}$ 
for the KK analysis. 
Then the background information is collected as 
the $\bthH$-dependent boundary conditions at one orbifold boundary. 
The nonlinear $\bthH$-dependences of the spectrum and the couplings 
are well seen in this approach. 
Since $\bthH$ corresponds to a vacuum expectation value (VEV) of 
the Higgs field, it is natural to promote it 
to the dynamical field in the above procedure. 
Namely we can see the nonlinear structure 
of the Higgs couplings manifestly 
by taking a gauge in which the fifth components of 
the {\it dynamical} gauge fields~$A_y$ are zero.\footnote{
This gauge is also useful to incorporate the 4D scalars 
coming from $A_y$ into the effective action 
in the holographic approach~\cite{Panico}. 
} 
Interestingly the resultant effective actions have 
similar forms to those of 4D models in which 
the Higgs bosons are provided as 
the pseudo Nambu-Goldstone bosons just like the models 
proposed in Ref.~\cite{littleHiggs}. 
This is consistent with the holographic interpretation. 

The paper is organized as follows. 
In the next section, we explain our method in detail 
to derive the effective theory of the GHU models 
in the warped spacetime. 
Then we give some comments on the resultant 
effective action. 
In Sec.~\ref{specific}, we apply our method 
to specific models and investigate them 
from the viewpoint of the effective theory. 
Especially we see the nonlinear structure 
of the Higgs sector explicitly. 
Sec.~\ref{summary} is devoted to the summary. 
The notations used in this paper are 
collected in Appendix~\ref{notation}. 
A complemental calculation to Sec.~\ref{int_terms} 
is shown in Appendix~\ref{cal_WWHH}.

\section{Derivation of 4D effective theory} \label{derivation_S4}
\subsection{Setup}
We consider a gauge theory with a gauge group~$\bG$ 
in the warped 5D spacetime. 
The fifth dimension is compactified on an orbifold~$S^1/Z_2$ with a radius~$R$. 
We use, throughout the paper, $M,N,\cdots=0,1,2,3,4$ for the 5D curved indices, 
$A,B,\cdots=0,1,2,3,4$ for the 5D flat indices in tetrads, 
and $\mu,\nu,\cdots=0,1,2,3$ for 4D indices.\footnote{
As the background geometry preserves 4D Poincar\'{e} invariance, 
the curved 4D indices are not discriminated from the flat ones. } 
The background metric is given by~\cite{RS}
\be
 ds^2 = G_{MN}dx^M dx^N = e^{-2\sgm(y)}\eta_{\mu\nu}dx^\mu dx^\nu+dy^2, 
 \label{metric}
\ee
where $\eta_{\mu\nu}=\diag(-1,1,1,1)$, $\sgm(y)=\sgm(y+2\pi R)$, and 
$\sgm(y)\equiv k\abs{y}$ for $\abs{y}\leq \pi R$. 
The cosmological constant in the bulk 5D spacetime is given by $\Lmd=-k^2$. 
The points~$(x^\mu,-y)$ and $(x^\mu,y+2\pi R)$ are identified with $(x^\mu,y)$. 
Thus the spacetime is equivalent to the interval in the fifth dimension 
with two boundaries at $y=0$ and $y=\pi R$, which we refer to as 
the UV brane and the IR brane, respectively. 

The 5D theory contains gauge fields~$A_M$ and 
a matter fermion field~$\Psi$. 
The former is decomposed as 
\be
 A_M = A^I_M T^I, 
\ee
where $T^I$ are the generators of $\bG$, which are normalized as 
\be
 \tr(T^IT^J) = \frac{1}{2}\dlt^{IJ}. 
\ee
The 5D action is 
\bea
 S_5 \eql \int\dr^5x\;\sqrt{-G}\sbk{-\frac{1}{2}\tr\brkt{F_{MN}F^{MN}}
 +i\bar{\Psi}\Gm^N\cD_N\Psi-iM\vep\bar{\Psi}\Psi}, 
 \label{5Daction}
\eea
where $G\equiv\det(G_{MN})$ and $\Gm^N\equiv e_A^{\;\;N}\Gm^A$. 
The integral region for $y$ is $\sbk{0,\pi R}$. 
The 5D $\gm$-matrices~$\Gm^A$ are related to the 4D ones~$\gm^\mu$ 
by $\Gm^\mu=\gm^\mu$ and $\Gm^4=\gm_5$ which is the 4D chiral operator. 
Note that $\Gm^0$ is anti-Hermitian in our notation. 
(See (\ref{Gm_rep}) in Appendix~\ref{notation}.) 
Since the operator~$\bar{\Psi}\Psi$ is anti-Hermitian and $Z_2$-odd, 
we need the factor~$i$ and the periodic 
sign function~$\vep(y)=\sgm'(y)/k$ satisfying~$\vep(y)=\pm 1$ 
in (\ref{5Daction}). 
The field strength and the covariant derivative are defined by 
\bea
 F_{MN} \defa \der_M A_N-\der_N A_M-ig_A\sbk{A_M,A_N}, \nonumber\\
 \cD_M\Psi \defa \brkt{\der_M-\frac{1}{4}\omg_M^{\;\;AB}\Gm_{AB}
 -ig_A A_M}\Psi, 
\eea
where $g_A$ is the 5D gauge coupling, 
and $\Gm^{AB}\equiv\frac{1}{2}\sbk{\Gm^A,\Gm^B}$. 
The spin connection 1-form~$\omg^{AB}=\omg_M^{\;\;AB}dx^M$ determined 
from the metric~(\ref{metric}) is 
\be
 \omg^{\nu 4} = -\frac{d\sgm}{dy}e^{-\sgm}dx^\nu, \;\;\;\;\;
 \mbox{other components} = 0. 
\ee

The orbifold $Z_2$-parity transformations around $y_0=0$ and $y_\pi=\pi R$, 
which preserve the orbifold structure, are written as 
\bea
 \vct{A_\mu}{A_y}(x,y_j-y) \eql P_j\vct{A_\mu}{-A_y}(x,y_j+y)P_j^{-1}, 
 \nonumber\\
 \Psi(x,y_j-y) \eql \eta_jP_j\gm_5\Psi(x,y_j+y), 
 \label{orbifold_cond}
\eea
where $\eta_j=\pm 1$ ($j=0,\pi$). 
The constant matrices~$P_j$ belong to the group~$\bG$ and satisfy $P_j^2=1$. 
The gauge group~$\bG$ is generically broken to a subgroup~$\bH$ 
due to the above orbifold conditions. 
The unbroken gauge group~$\bH$ depends on the choice of $P_j$ ($j=0,\pi$). 
The generators~$T^I$ are then classified into two parts, \ie, 
$T^a$ and $T^{\ha}$, which are the generators of $\bH$ and $\bG/\bH$, 
respectively. 
In this paper we take the same $Z_2$-parity assignment 
at both boundaries, for simplicity. 
This assumption can be easily relaxed in the following derivation 
of the 4D effective theory. 

The fermion field~$\Psi$ belongs to an irreducible representation 
of the full gauge group~$\bG$. 
It can be decomposed into components 
each of which belongs to an irreducible representation of $\bH$. 
For simplicity, we assume that $\Psi$ is decomposed into 
two such components, \ie,  
\be
 \Psi = \vct{q}{Q},  \label{decomp_Psi}
\ee
where $q$ and $Q$ belong to irreducible representations of $\bH$ 
and have opposite $Z_2$-parities. 
In the case of $\bG=SU(3)$ and $\bH=SU(2)\times U(1)$, 
for example, an $SU(3)$ triplet~$\Psi$ is decomposed 
into a doublet~$q$ and a singlet~$Q$ for the unbroken $SU(2)$. 
Extension to the cases that $\Psi$ is decomposed into 
more than two components is straightforward. 

The $Z_2$-parities for the fields are collected in Table~\ref{Z2_parity}. 
\begin{table}[t, b]
\begin{center}
\begin{tabular}{|c|c|c|c||c|c|c|c|}
 \hline \rule[-2mm]{0mm}{7mm} $A^a_\mu$ & $A^{\ha}_\mu$ & 
 $A^a_y$ & $A^{\ha}_y$  & $q_L$ & $Q_L$ & $q_R$ & $Q_R$ \\ \hline 
 $+$ & $-$ & $-$ & $+$ & $+$ & $-$ & $-$ & $+$ \\ \hline 
\end{tabular}
\end{center}
\caption{The $Z_2$-parities of the gauge and fermion fields. 
We take the same parity assignment at both boundaries.}  
\label{Z2_parity}
\end{table}
The fields which are even (odd) under the orbifold parity 
at $y=y_j$ obey the Neumann (Dirichlet) boundary conditions there 
if there are no additional dynamics on the boundary~$y=y_j$.

\subsection{Gauging away of $A_y$} \label{gauge_away}
We can always gauge $A^a_y$ away 
without changing the boundary conditions listed in Table~\ref{Z2_parity}. 
Namely $A^a_y$ are not physical degrees of freedom. 
On the other hand, $A^{\ha}_y$ cannot be completely gauged away 
because corresponding gauge transformation mixes components 
with different $Z_2$-parities. 
This means that $A^{\ha}_y$ contain physical modes. 
However we can formally perform the following gauge transformation 
to remove the whole~$A_y$ from the bulk. 
\be
 \tl{A}_M = \Omg A_M \Omg^{-1}-\frac{i}{g_A}(\der_M \Omg)\Omg^{-1}, 
\ee
where 
\be
 \Omg(x,y) \equiv \cP\exp\brc{ig_A\int_y^{\pi R}\dr y'\;A_y(x,y')}. 
\ee
The simbol~$\cP$ stands for the path ordering operator from left to right. 
\ignore{
$\Omg$ is a solution of 
\be
 \Omg^{-1}\der_y\Omg = -ig_A A_y^{\ha}T^{\ha}. 
 \label{eq_for_Omg}
\ee
}
This transformation must be discontinuous at $y=0$ 
to maintain the $Z_2$-parities of the fields. 
If we parametrize $\Omg$ as 
\be
 \Omg(x,y)=\exp\brc{i\vph^a(x,y)T^a}\exp\brc{i\tht^{\ha}(x,y)T^{\ha}}, 
\ee
the transformation parameters~$\tht^{\ha}(x,y)$ are odd under 
the $Z_2$-parity at both boundaries while $\vph^a(x,y)$ are even. 
Since $\Omg(x,y)$ satisfies that 
\be
 \Omg(x,\pi R) = 1, \label{ini_cond}
\ee
\ie, $\vph^a(x,\pi R)=\tht^{\ha}(x,\pi R)=0$, 
all the transformation parameters~$(\vph^a,\tht^{\ha})$ are continuous 
at the IR brane. 
On the other hand, the parameters~$\tht^{\ha}(x,y)$ become discontinuous 
at the UV brane, \ie, 
\be
 \sbk{\tht^{\ha}(x,y)}_{y=-\ep}^{y=\ep} = 2\tht^{\ha}(x,\ep) \neq 0,  
 \label{tht_gap}
\ee
where $\ep$ is a positive infinitesimal. 
These gaps~$\tht^{\ha}(x,\ep)$ at $y=0$, 
which are equal to the Wilson line phases by definition,  
are the physical degrees of freedom, 
and are identified with the Higgs fields as we will see below. 
The parameters~$\vph^a(x,y)$ are continuous also at the UV brane, 
and can be absorbed by a gauge transformation for 
the residual gauge symmetry~$\bH$. 

Now the action~(\ref{5Daction}) becomes 
\bea
 S_5 \eql S^{\rm gauge}_5+S^{\rm fermi}_5, \nonumber\\
 S^{\rm gauge}_5 \eql \int\dr^5x\;
 \sbk{-\frac{1}{2}\tr\brc{\tl{F}_{\mu\nu}\tl{F}^{\mu\nu}
 +2e^{-2\sgm}\brkt{\der_y\tl{A}_\mu\der_y\tl{A}^\mu}}}, 
 \nonumber\\
 S^{\rm fermi}_5 \eql \int\dr^5x\;
 i\brc{\bar{\tl{\Psi}}\brkt{e^\sgm\gm^\mu\tl{\cD}_\mu+\gm_5\der_y-M\vep}
 \tl{\Psi}}, \label{5Daction2}
\eea
where 
\bea
 \tl{\Psi} \eql \vct{\tl{q}}{\tl{Q}} \equiv e^{-2\sgm}\Omg\Psi, 
 \nonumber\\
 \tl{\cD}_\mu\tl{\Psi} \defa \brkt{\der_\mu-ig_A\tl{A}_\mu}\tl{\Psi}.  
\eea
\ignore{
In the following we take the integral interval for $y$ in the 5D action 
as $\sbk{\ep,\pi R-\ep}$ in order to avoid a careful treatment of 
the orbifold singularity. 
This is possible because we do not introduce any dynamics 
on the boundaries. 
Thus $\vep(y)=1$. }
The boundary conditions for $(\tl{A}_\mu,\tl{\Psi})$ at $y=\pi R$ 
are the same as those of $(A_\mu,e^{-2\sgm(y)}\Psi)$ 
because $\Omg(x,\pi R)=1$. 
Namely they obey the conditions: 
\bea
 \der_y\tl{A}^a_\mu \eql 0, \;\;\;\;\;
 \tl{A}^{\ha}_\mu = 0, \label{IRbdcd:gauge} \\
 (\der_y+M)\tl{q}_L \eql 0, \;\;\;\;\;
 \tl{Q}_L = 0, \nonumber\\
 \tl{q}_R \eql 0, \;\;\;\;\;
 (\der_y-M)\tl{Q}_R = 0,  \label{IRbdcd:fermion}
\eea
at $y=\pi R$. 
On the other hand, the boundary conditions at $y=0$ are 
neither Neumann nor Dirichlet conditions any more  
due to the appearance of $\tht^{\ha}(x,\ep)$. 
In the rest of this paper we take the integration region for $y$ 
as $\sbk{\ep,\pi R}$ in order to avoid 
the discontinuity at $y=0$. 
Then $\sgm(y)=ky$ and $\vep(y)=1$ in the following.

\subsection{Gauge sector} \label{gauge_sector}
The equations of motion for the gauge fields~$\tl{A}^I_\mu$ are obtained 
from (\ref{5Daction2}) as  
\be
 \sbk{\cD_\nu\tl{F}^{\mu\nu}}^I-\der_y\brkt{e^{-2\sgm}\der_y\tl{A}^{I\mu}}
 +e^\sgm g_A\bar{\tl{\Psi}}\gm^\mu T^I\tl{\Psi} = 0. 
 \label{EOM:gauge}
\ee
The symbol~$\sbk{\cdots}^I$ denotes the $I$-component 
in the decomposition by the generators~$T^I$, \ie, 
\be
 \sbk{\cC}^I \equiv 2\tr\brkt{T^I \cC} = \cC^I, 
\ee
for some matrix~$\cC=\cC^JT^J$. 
From the 4D point of view, the second term in (\ref{EOM:gauge}) 
corresponds to the mass term. 
We focus on the modes that are much lighter than 
the KK mass scale~$\mKK\equiv k\pi/(e^{k\pi R}-1)$ in the following. 
They can be regarded as massless modes in the first approximation. 
Thus we impose the following ``massless condition'' to extract 
the zero-modes from the 5D fields. 
\be
 \der_y\brkt{e^{-2\sgm}\der_y\tl{A}^I_\mu} = 0. 
 \label{ms_cond:gauge}
\ee
Solving this with the conditions~(\ref{IRbdcd:gauge}), the $y$-dependences 
of $(\tl{A}^a_\mu,\tl{A}^{\ha}_\mu)$ are completely determined as 
\bea
 \tl{A}^a_\mu(x,y) \eql \tl{A}^a_\mu(x,\ep), \nonumber\\
 \tl{A}^{\ha}_\mu(x,y) \eql \frac{e^{2k\pi R}-e^{2\sgm(y)}}{e^{2k\pi R}-1}
 \tl{A}^{\ha}_\mu(x,\ep). 
 \label{gen_sol:gauge}
\eea
Substituting these into $S^{\rm gauge}_5$ in (\ref{5Daction2}) 
and performing the $y$-integral, 
we obtain the 4D action. 
\be
 S_4^{\rm gauge} \simeq \int\dr^4x\;\brc{-\frac{\pi R}{2}
 \tr\brkt{\tl{F}_{\mu\nu}\tl{F}^{\mu\nu}}
 -ke^{-2k\pi R}\tl{A}^{\ha}_{\mu}\tl{A}^{\ha\mu}}_{y=\ep}. 
 \label{S4:gauge}
\ee
Here we have neglected corrections suppressed by a factor of 
$(k\pi R)^{-1}$, which is about a few percent 
when $e^{k\pi R}=\cO(10^{15})$. 
Note that each component of $\tl{A}_\mu(x,\ep)$ is expressed by 
$A^a_\mu(x,\ep)$, $\vph^a(x,\ep)$ and $\tht^{\ha}(x,\ep)$ 
since $A^{\ha}_\mu(x,0) = 0$. 
As mentioned in the previous subsection, 
the $\vph^a(x,\ep)$-dependence can be removed 
by a 4D gauge transformation for $\bH$. 
In the following the 4D fields~$\tl{A}^I_\mu(x,\ep)$ 
are understood as the fields after this gauge transformation. 
Then they are expressed in terms of $A^a_\mu(x,\ep)$ 
and $\tht^{\ha}(x,\ep)$. 
For example, 
\bea
 \tl{A}^{\ha}_\mu(x,\ep) \eql 
 \sbk{\Omg_0 A_\mu(x,\ep)\Omg_0^{-1}
 -\frac{i}{g_A}\der_\mu\Omg_0\Omg_0^{-1}}^{\ha} \nonumber\\
 \eql \frac{1}{g_A}\brc{
 \der_\mu\tht^{\ha}(x,\ep)
 +g_AC_{\ha b\hat{c}}A^b_\mu(x,\ep)\tht^{\hat{c}}(x,\ep)+\cdots},
 \label{tlA^ha}
\eea
where $C_{IJK}$ are the structure constants of $\bG$ 
defined in (\ref{def:C_IJK}), and 
\be
 \Omg_0(x) \equiv \exp\brc{i\tht^{\ha}(x,\ep)T^{\ha}}. 
\ee
The ellipsis in the second line of (\ref{tlA^ha}) 
denotes higher order terms for $\tht^{\ha}(x,\ep)$. 
Thus the second term in (\ref{S4:gauge}) corresponds to the kinetic terms 
for $\tht^{\ha}(x,\ep)$. 
On the other hand, the $\tht^{\ha}$-dependence of the first term 
in (\ref{S4:gauge}) is cancelled, \ie, 
\be
 \tr\brkt{\tl{F}_{\mu\nu}\tl{F}^{\mu\nu}}_{y=\ep}
 = \tr\brkt{F_{\mu\nu}F^{\mu\nu}}_{y=\ep} 
 = \frac{1}{2}\brkt{F^a_{\mu\nu}F^{a\mu\nu}}_{y=\ep}, 
\ee
Therefore we redefine the fields as 
\bea
 \cA^a_\mu(x) \defa \sqrt{\pi R}A^a_\mu(x,\ep), \nonumber\\
 H^{\ha}(x) \defa \frac{\sqrt{2k}e^{-k\pi R}}{g_A}\tht^{\ha}(x,\ep), 
 \label{def:cA}
\eea
so that the fields are canonically normalized. 
The resultant 4D action is 
\be
 S_4^{\rm gauge} \simeq \int\dr^4x\;\brc{
 -\frac{1}{4}\cF^a_{\mu\nu}\cF^{a\mu\nu}
 -\frac{1}{2}\tl{\cD}^{(4)}_\mu H^{\ha}
 \tl{\cD}^{(4)\mu} H^{\ha}},   \label{rtS4:gauge}
\ee
where 
\bea
 \cF^a_{\mu\nu} \defa \der_\mu\cA^a_\nu-\der_\nu\cA^a_\mu
 +\bar{g}_AC_{abc}\cA^b_\mu\cA^c_\nu, \nonumber\\
 \tl{\cD}^{(4)}_\mu H^{\ha} \defa \frac{\sqrt{2k}e^{-k\pi R}}{\sqrt{\pi R}}
 \sbk{\Omg_0(\cA^a_\mu T^a)\Omg_0^{-1}
 -\frac{i}{\bar{g}_A}\der_\mu\Omg_0\Omg_0^{-1}}^{\ha}  \nonumber\\
 \eql \der_\mu H^{\ha}+\bar{g}_AC_{\ha b\hat{c}}\cA^b_\mu H^{\hat{c}}
 +\cO(H^2).  \label{def_DthH}
\eea
Here a dimensionless constant~$\bar{g}_A \equiv g_A/\sqrt{\pi R}$ is 
the 4D gauge coupling, and $\Omg_0$ is expressed in terms of $H^{\ha}$ 
as 
\bea
 \Omg_0(x) \eql \exp\brc{\frac{i}{\fH}H^{\ha}(x)T^{\ha}}, \nonumber\\
 \fH \defa \frac{\sqrt{2k}e^{-k\pi R}}{g_A}. 
 \label{def:Omg0}
\eea

\subsection{Fermion sector}
The equation of motion for the fermion field~$\tl{\Psi}$ is 
\be
 \brc{\gm^\mu\tl{\cD}_\mu+e^{-\sgm}\brkt{\gm_5\der_y-M}}\tl{\Psi} = 0. 
 \label{EOM:fermion}
\ee
Multiplying the differential operator:~$\brc{\gm^\nu\tl{\cD}_\nu+e^{-\sgm}
\brkt{\gm_5\der_y+M}}$ from the left, we obtain 
\be
 \tl{\cD}_\mu\tl{\cD}^\mu\tl{\Psi}+e^{-2\sgm}
 \brc{\der_y^2-k\brkt{\der_y-\gm_5 M}-M^2}\tl{\Psi} = 0. 
 \label{EOM2:fermion}
\ee
The second term corresponds to the mass term from the 4D point of view. 
Then the ``massless conditions'' are written as 
\be
 \brc{\der_y^2-k(\der_y-\gm_5 M)-M^2}\vct{\tl{q}}{\tl{Q}} = 0. 
 \label{ms_cond:fermion}
\ee
Solving these equations with the boundary conditions in (\ref{IRbdcd:fermion}), 
the $y$-dependence of the 5D fermion fields are completely determined as follows. 
\bea
 \tl{q}_L(x,y) \eql e^{-c\sgm(y)}\tl{q}_L(x,\ep), \nonumber\\
 \tl{Q}_L(x,y) \eql e^{-c\sgm(y)}\brkt{
 \frac{e^{(k+2M)\pi R}-e^{(1+2c)\sgm(y)}}{e^{(k+2M)\pi R}-1}}\tl{Q}_L(x,\ep), 
 \nonumber\\
 \tl{q}_R(x,y) \eql e^{c\sgm(y)}\brkt{
 \frac{e^{(k-2M)\pi R}-e^{(1-2c)\sgm(y)}}{e^{(k-2M)\pi R}-1}}\tl{q}_R(x,\ep), 
 \nonumber\\
 \tl{Q}_R(x,y) \eql e^{c\sgm(y)}\tl{Q}_R(x,\ep), 
 \label{gen_sol:fermion}
\eea
where $c\equiv M/k$. 
Each component of $\tl{\Psi}(x,\ep)$ is expressed by 
$q_L(x,\ep)$, $Q_R(x,\ep)$, $\vph^a(x,\ep)$ and $\tht^{\ha}(x,\ep)$ 
since $Q_L(x,0)=q_R(x,0)=0$. 
As mentioned in the previous subsection, $\vph^a$ can be 
removed by the 4D gauge transformation for $\bH$. 
So $\tl{\Psi}(x,\ep)$ is understood as the 4D field 
after this transformation. 
Namely, 
\bea
 \vct{\tl{q}_L}{\tl{Q}_L}(x,\ep) \eql \Omg_0\vct{q_L}{0}(x,\ep), \nonumber\\
 \vct{\tl{q}_R}{\tl{Q}_R}(x,\ep) \eql \Omg_0\vct{0}{Q_R}(x,\ep). 
 \label{qQ_rewrite}
\eea
Making use of the following equations followed from (\ref{gen_sol:fermion}), 
\be
 (\der_y+M)\tl{q}_L = (\der_y-M)\tl{Q}_R = 0, 
\ee
the fermionic part of the 5D action~$S^{\rm fermi}_5$ in (\ref{5Daction2}) 
can be rewritten as 
\bea
 S_5^{\rm fermi} \eql \int\dr^5x\;i\brc{e^{\sgm}\bar{\tl{\Psi}}\gm^\mu\tl{\cD}_\mu
 \tl{\Psi}+\der_y\brkt{\bar{\tl{q}}_L\tl{q}_R-\bar{\tl{Q}}_R\tl{Q}_L}} 
 \nonumber\\
 \eql \int\dr^5x\;ie^{\sgm}\bar{\tl{\Psi}}\gm^\mu\tl{\cD}_\mu\tl{\Psi}
 +\int\dr^4x\;i\brkt{-\bar{\tl{q}}_L\tl{q}_R+\bar{\tl{Q}}_R\tl{Q}_L}_{y=\ep}. 
 \label{S5:fermion2}
\eea
The surface terms on the IR brane vanish 
due to the boundary conditions~(\ref{IRbdcd:fermion}). 

In the following we assume that $\abs{c}\geq 1/2$.

\subsubsection{case of $c\geq 1/2$}
The solution~(\ref{gen_sol:fermion}) is approximated as 
\bea
 \tl{q}_L(x,y) \eql e^{-c\sgm(y)}\tl{q}_L(x,\ep), \;\;\;\;\;
 \tl{Q}_L(x,y) \simeq e^{-c\sgm(y)}\tl{Q}_L(x,\ep), \nonumber\\
 \tl{q}_R(x,y) \sqa e^{(1-c)\sgm(y)}\tl{q}_R(x,\ep), 
 \;\;\;\;\;
 \tl{Q}_R(x,y) = e^{c\sgm(y)}\tl{Q}_R(x,\ep).  
\eea
Substituting these into (\ref{S5:fermion2}) 
and performing the $y$-integral, we obtain the 4D action, 
\bea
 S^{\rm fermi}_4 \sqa \int\dr^4x\;i\left\{\frac{1}{2M-k}
 \bar{\tl{\Psi}}_L\gm^\mu\tl{\cD}_\mu\tl{\Psi}_L 
 +\frac{e^{(k+2M)\pi R}}{k+2M}(0,\bar{\tl{Q}}_R)\gm^\mu\tl{\cD}_\mu
 \vct{0}{\tl{Q}_R} \right. \nonumber\\
 &&\hspace{15mm}\left. 
 -\bar{\tl{q}}_L\tl{q}_R+\bar{\tl{Q}}_R\tl{Q}_L\right\}_{y=\ep}.  
 \label{S4:fermion}
\eea
Since $e^{(1-c)\sgm(y)}\ll e^{c\sgm(y)}$ in most part of the bulk, 
the contribution of $\tl{q}_R$ in the $y$-integral in (\ref{S5:fermion2}) 
is negligible.
We have dropped corrections suppressed by a factor of $(k\pi R)^{-1}$ 
coming from the integral of terms involving $\tl{A}^{\ha}_\mu$ 
in $\tl{\cD}_\mu$. 

We can see from (\ref{qQ_rewrite}) that 
the $\tht^{\ha}$-dependence of the first term in (\ref{S4:fermion}) 
is cancelled, \ie,  
\be
 \brkt{\bar{\tl{\Psi}}_L\gm^\mu\tl{\cD}_\mu\tl{\Psi}_L}_{y=\ep}
 = \brkt{(\bar{q}_L,0)\gm^\mu\hat{\cD}_\mu \vct{q_L}{0}}_{y=\ep}, 
\ee 
where
\be
 \hat{\cD}_\mu \equiv \der_\mu-ig_A A_\mu. 
\ee
On the other hand, such cancellation does not occur 
for the right-handed components because the contribution of $\tl{q}_R$ 
in the kinetic term is negligible.  

Here we decompose $\Omg_0$ into four parts so that the relation: 
$\tl{\Psi}(x,\ep)=\Omg_0\Psi(x,\ep)$ is rewritten as
\be
 \vct{\tl{q}(x,\ep)}{\tl{Q}(x,\ep)} = 
 \mtrx{\Omg_0^{qq}}{\Omg_0^{qQ}}{\Omg_0^{Qq}}{\Omg_0^{QQ}}
 \vct{q(x,\ep)}{Q(x,\ep)}. 
\ee
Then (\ref{S4:fermion}) becomes 
\bea
 S^{\rm fermi}_4 \sqa \int\dr^4x\;i\left\{
 \frac{1}{2M-k}(\bar{q}_L,0)\gm^\mu\hat{\cD}_\mu \vct{q_L}{0}
 +\frac{e^{(k+2M)\pi R}}{k+2M}(0,\bar{Q}_R(\Omg_0^{QQ})^\dagger)
 \gm^\mu\tl{\cD}_\mu\vct{0}{\Omg_0^{QQ}Q_R} \right. \nonumber\\
 &&\hspace{15mm}\left. 
 -\bar{q}_L(\Omg_0^{qq})^\dagger \Omg_0^{qQ}Q_R
 +\bar{Q}_R(\Omg_0^{QQ})^\dagger \Omg_0^{Qq}q_L \right\}_{y=\ep}. 
\eea
To normalize the fields canonically, we redefine them as 
\bea
 \psi_L(x) \defa \frac{q_L(x,\ep)}{\sqrt{2M-k}}, \label{def:psiL}\\
 \chi_R(x) \defa i\brkt{\frac{e^{(k+2M)\pi R}}{k+2M}}^{1/2}\Omg_0^{QQ}Q_R(x,\ep) 
 \nonumber\\
 \eql \frac{ie^{-\frac{3}{2}k\pi R}}{\sqrt{k+2M}}Q_R(x,\pi R). 
 \label{def:chiR}
\eea
Then the above action is rewritten as 
\bea
 S^{\rm fermi}_4 \sqa \int\dr^4x\;i\left\{
 \bar{\psi}_L\gm^\mu\cD^{(4)}_\mu\psi_L
 +\bar{\chi}_R\gm^\mu\tl{\cD}^{(4)}_\mu\chi_R \right.
 \nonumber\\
 &&\hspace{15mm}\left.
 -i\brkt{\frac{4M^2-k^2}{e^{(2M+k)\pi R}}}^{1/2}\brkt{
 \bar{\psi}_L(\Omg_0^{Qq})^\dagger\chi_R-\bar{\chi}_R\Omg_0^{Qq}\psi_L}
 \right\}.  \label{rtS4:fermion}
\eea
Here we have used the relation followed from the unitarity condition of $\Omg_0$:
\be
 (\Omg_0^{qq})^\dagger\Omg_0^{qQ} = -(\Omg_0^{Qq})^\dagger\Omg_0^{QQ}. 
 \label{Omg0:unitarity}
\ee
The covariant derivatives~$\cD^{(4)}_\mu$ and $\tl{\cD}^{(4)}_\mu$ are defined by 
\bea
 \cD^{(4)}_\mu\psi_L \defa \brkt{\der_\mu-i\bar{g}_A\cA^a_\mu T^a}\psi_L, 
 \nonumber\\
 \tl{\cD}^{(4)}_\mu\chi_R \defa \brkt{\der_\mu
 -i\bar{g}_A\tl{\cA}^a_\mu T^a}\chi_R, 
 \label{def:cD4}
\eea
where
\be
 \tl{\cA}^a_\mu \equiv \sbk{\Omg_0(\cA^a_\mu T^a)\Omg_0^{-1}
 -\frac{i}{\bar{g}_A}\der_\mu\Omg_0\Omg_0^{-1}}^a  
 \label{def_tlcA}
\ee

\subsubsection{case of $c\leq -1/2$}
Now the solution~(\ref{gen_sol:fermion}) is approximated as 
\bea
 \tl{q}_L(x,y) \eql e^{-c\sgm(y)}\tl{q}_L(x,\ep), \;\;\;\;\;
 \tl{Q}_L(x,y) \simeq e^{(1+c)\sgm(y)}\tl{Q}_L(x,\ep), \nonumber\\
 \tl{q}_R(x,y) \sqa e^{c\sgm(y)}\tl{q}_R(x,\ep), 
 \;\;\;\;\;
 \tl{Q}_R(x,y) = e^{c\sgm(y)}\tl{Q}_R(x,\ep). 
\eea
Then $S^{\rm fermi}_4$ is calculated as
\bea
 S^{\rm fermi}_4 \sqa \int\dr^4x\;i\left\{
 \frac{e^{(k-2M)\pi R}}{k-2M}(\bar{\tl{q}}_L,0)\gm^\mu\tl{\cD}_\mu
 \vct{\tl{q}_L}{0}+\frac{1}{k+2M}\bar{\tl{\Psi}}_R\gm^\mu\tl{\cD}_\mu
 \tl{\Psi}_R \right.\nonumber\\
 &&\hspace{15mm}\left.
 -\bar{\tl{q}}_L\tl{q}_R+\bar{\tl{Q}}_R\tl{Q}_L\right\}_{y=\ep}. 
 \label{S4:fermion2}
\eea
Since $e^{-c\sgm(y)}\gg e^{(1+c)\sgm(y)}$ in most part of the bulk, 
the contribution of $\tl{Q}_L$ in the kinetic term is negligible. 
We have again dropped terms suppressed by a factor of $(k\pi R)^{-1}$. 

The $\tht^{\ha}$-dependence of the second term in 
(\ref{S4:fermion2}) is cancelled, \ie,  
\be
 \brc{\bar{\tl{\Psi}}_R\gm^\mu\tl{\cD}_\mu\tl{\Psi}_R}_{y=\ep}
 = \brc{(0,\bar{Q}_R)\gm^\mu\hat{\cD}_\mu\vct{0}{Q_R}}_{y=\ep}, 
\ee
while such cancellation does not occur 
for the left-handed components. 
Thus (\ref{S4:fermion2}) becomes 
\bea
 S^{\rm fermi}_4 \sqa \int\dr^4x\;i\left\{
 \frac{e^{(k-2M)\pi R}}{k-2M}(\bar{q}_L(\Omg_0^{qq})^\dagger,0)
 \gm^\mu\tl{\cD}_\mu\vct{\Omg_0^{qq}q_L}{0}
 +\frac{1}{k+2M}(0,\bar{Q}_R)\gm^\mu\hat{\cD}_\mu\vct{0}{Q_R} 
 \right. \nonumber\\
 &&\hspace{15mm}\left.
 -\bar{q}_L(\Omg_0^{qq})^\dagger\Omg_0^{qQ}Q_R
 +\bar{Q}_R(\Omg_0^{QQ})^\dagger\Omg_0^{Qq}q_L \right\}_{y=\ep}. 
\eea
To normalize the fields canonically, we redefine them as 
\bea
 \chi_L(x) \defa i\brkt{\frac{e^{(k-2M)\pi R}}{k-2M}}^{1/2}
 \Omg_0^{qq}q_L(x,\ep) \nonumber\\
 \eql \frac{ie^{-\frac{3}{2}k\pi R}}{\sqrt{k-2M}}q_L(x,\pi R), 
 \label{def:chiL}\\
 \psi_R(x) \defa \frac{Q_R(x,\ep)}{\sqrt{k+2M}}.  \label{def:psiR}
\eea
Then the above action is rewritten as 
\bea
 S^{\rm fermi}_4 \sqa \int\dr^4x\;i\left\{
 \bar{\chi}_L\gm^\mu\tl{\cD}^{(4)}_\mu\chi_L
 +\bar{\psi}_R\gm^\mu\cD^{(4)}_\mu\psi_R \right. \nonumber\\
 &&\hspace{15mm}\left. 
 -i\brkt{\frac{k^2-4M^2}{e^{(k-2M)\pi R}}}^{1/2}
 \brkt{\bar{\chi}_L\Omg_0^{qQ}\psi_R
 -\bar{\psi}_R(\Omg_0^{qQ})^\dagger\chi_L}\right\}. 
 \label{rtS4:fermion2}
\eea
Here we have used the Hermitian conjugate of (\ref{Omg0:unitarity}), 
\be
 (\Omg_0^{QQ})^\dagger\Omg_0^{Qq} = -(\Omg_0^{qQ})^\dagger\Omg_0^{qq}. 
\ee

\subsection{comments} \label{comments}
\subsubsection{Review of 4D action}
Let us review the derived 4D effective action. 
It consists of the Higgs fields~$H^{\ha}(x)$ 
(or the Wilson line phase~$\tht^{\ha}(x,\ep)$) 
and the 4D boundary values of the 5D fields at the UV brane. 
Note that the gauge fields~$\tl{\cA}^a_\mu$ 
that appear in the covariant derivatives~$\tl{\cD}^{(4)}_\mu\chi_{R,L}$ 
are dressed by the Higgs fields. 
Such ``dressed gauge fields''~$\tl{\cA}^a_\mu$ 
and the covariant derivatives of 
the Higgs fields~$\tl{\cD}_\mu^{(4)}H^{\ha}$ 
can be read off from the gauged Maurer Cartan 1-form~$\alp_\mu$ 
as follows.  
(See (\ref{def_DthH}) and (\ref{def_tlcA}).)
\bea
 i\alp_\mu \defa \Omg_0\brkt{\bar{g}_A\cA^a_\mu T^a}\Omg_0^{-1}
 -i\der_\mu\Omg_0\Omg_0^{-1} \nonumber\\
 \eql \bar{g}_A\tl{A}^a_\mu T^a
 +\frac{1}{\fH}\tl{\cD}^{(4)}_\mu H^{\ha}T^{\ha}. 
 \label{def:alp_mu}
\eea
Here $\Omg_0$ is an element of $\bG/\bH$ parametrized by the Higgs fields as 
(\ref{def:Omg0}). 
Using these ingredients we can construct 
the effective action~$S_4=S^{\rm gauge}_4+S^{\rm fermi}_4$ 
by the formulae~(\ref{rtS4:gauge}) and (\ref{rtS4:fermion}) 
(or (\ref{rtS4:fermion2})). 

\subsubsection{Analogy to PNG Higgs models and holographic interpretation}
Obviously the above construction is nothing but 
the nonlinear realization of $\bG/\bH$. 
This suggests an analogy between 
5D GHU model in the warped spacetime 
and 4D model where the Higgs bosons are realized as 
the pseudo Nambu-Goldstone (PNG) bosons 
just like the models in Ref.~\cite{littleHiggs}. 
This analogy has been discussed in detail in Ref.~\cite{holograph}. 
The Higgs fields~$H^{\ha}$ and the constant~$\fH$ 
correspond to the PNG bosons and their decay constant, respectively.  
The large gauge invariance of $S_4$ 
(or the periodicity of $H^{\ha}$) is manifest 
since $H^{\ha}$ are the coordinates on the compact manifold~$\bG/\bH$. 
If we formally set $\cA^a_\mu$ and $\psi_{L,R}$ to zero, 
the effective action~$S_4$ is invariant under the following 
nonlinear $\bG$-transformations. 
\be
 H^{\ha} \to H^{'\ha}, \;\;\;\;\;
 \chi_{R,L} \to h \chi_{R,L}, 
\ee
where $H^{'\ha}$ and $h\in\bH$ are defined as 
\bea
 \xi\Omg_0^{-1}(H^{\ha}) = \Omg_0^{-1}(H^{'\ha})h, 
\eea
for an arbitrary group element~$\xi\in\bG$. 
Thus $h$ depends not only on $\xi$ but also on $H^{\ha}$. 
Under the above transformations, 
it follows from (\ref{def:cD4}) and (\ref{def:alp_mu}) that 
\bea
 \tl{\cA}^a_\mu T^a \toa h(\tl{\cA}^a_\mu T^a)h^{-1}
 -\frac{i}{\bar{g}_A}\der_\mu hh^{-1}, \nonumber\\
 \tl{\cD}^{(4)}_\mu H^{\ha}T^{\ha} \toa 
 h(\tl{\cD}^{(4)}_\mu H^{\ha}T^{\ha})h^{-1}, \nonumber\\
 \tl{\cD}^{(4)}_\mu\chi_{R,L} \toa h\tl{\cD}^{(4)}_\mu\chi_{R,L}, 
\eea
Here $\tl{\cA}^a_\mu$ are purely made of $H^{\ha}$. 
The invariance under these transformations 
ensures the masslessness of $H^{\ha}$. 
Namely they can be identified as exact NG bosons in this case. 
Now we turn on $\cA^a_\mu$ and $\psi_{L,R}$ in the effective action. 
Then the above $\bG$ invariance is broken to $\bH$ explicitly 
so that the NG bosons~$H^{\ha}$ become PNG bosons.\footnote{
The Higgs fields~$H^{\ha}$ acquire nonzero masses at quantum level. } 
Recall that both $\chi_{R,L}$ and $H^{\ha}$ correspond to 
zero-modes of $\Psi$ and $A_y^{\ha}$ localized near the IR brane 
in the KK analysis. 
\ignore{
is a zero-mode of $\Psi$ localized near the IR brane, and 
$H^{\ha}$ originate from zero-modes of $A_y^{\ha}$, 
which are also localized near the IR brane in the KK analysis. }
On the other hand $\psi_{L,R}$ and $\cA^a_\mu$ correspond to 
zero-modes of $\Psi$ and $A_\mu^a$ that are 
localized near the UV brane and spread over the bulk, respectively. 
Hence only the modes localized near the IR brane respect 
the $\bG$ symmetry which is realized nonlinearly. 

The above features are consistent with 
the holographic interpretation~\cite{LPR,holograph,AC,Panico}, 
which is based on a conjecture that 5D theories on the warped spacetime 
are dual to 4D theories with a strongly interacting sector. 
In this interpretation, the 5D bulk corresponds to some strongly coupled 
conformal sector in a 4D theory, and the UV and the IR branes correspond to 
the UV cutoff scale~$\Lmd_{\rm UV}$ 
and the spontaneous breakdown of the conformal symmetry at 
$\Lmd_{\rm IR}$, respectively. 
Due to the conformal symmetry breaking, 
there appears a mass gap in the theory 
and the CFT spectrum is discretized. 
Namely bound states appear, 
which are identified with the modes localized near the IR brane 
in the 5D picture. 
There is a massless bound state~$\chi_R$ or $\chi_L$ among such modes 
depending on the parameter~$c=M/k$, which corresponds to 
the dimension of a CFT operator relevant to $\chi_{R,L}$. 
The gauge symmetry~$\bG$ in the 5D theory is identified with 
a global symmetry in the conformal sector of the 4D theory, 
and is spontaneously broken to the subgroup~$\bH$ at $\Lmd_{\rm IR}$. 
The unbroken symmetry~$\bH$ is gauged by $\cA^a_\mu$ that are 
external to the conformal sector. 
The elementary fields~$\cA^a_\mu$ and $\psi_{L,R}$, which are coupled 
to the conformal sector at $\Lmd_{\rm UV}$, are provided by 
the boundary values of the 5D fields~$A^a_\mu$ and $\Psi$ 
at the UV brane. 
This is consistent with (\ref{def:cA}) and (\ref{def:psiL}) 
(or (\ref{def:psiR})).\footnote{
Precisely speaking, the UV boundary values of the 5D fields contain 
contributions from the KK modes which we have neglected. 
However they are exponentially suppressed because the KK modes 
are localized near the IR brane in the warped spacetime. } 
In this holographic interpretation, the facts mentioned in the previous 
paragraph are understood as follows. 
If we turn off the external fields~$\cA^a_\mu$ and $\psi_{L,R}$, 
the global symmetry~$\bG$ becomes exact and thus the NG modes~$H^{\ha}$  
associated with $\bG/\bH$ are massless. 
The effective action consists of the bound states~$H^{\ha}$ and $\chi_{R,L}$, 
and has an invariance under the nonlinear $\bG$-transformation. 
The gauge connections~$\tl{\cA}^a_\mu$ in $\tl{\cD}^{(4)}_\mu\chi_{R,L}$ 
are purely made of the NG bosons~$H^{\ha}$, \ie, the CFT bound states.  
After including the external fields~~$\cA^a_\mu$ and $\psi_{L,R}$, 
$\bG$ is broken to $\bH$ explicitly so that $H^{\ha}$ become PNG bosons. 
The connection~$\tl{\cA}^a_\mu$ are now 
the mixing states between the elementary states~$\cA^a_\mu$ and 
the CFT bound states. 

The symmetry structure mentioned above can also be seen 
in Ref.~\cite{Panico} where the 4D effective action is derived 
in the holographic procedure. 
The holographic procedure is useful to calculate 
the effective potential of the Higgs fields or 
the electroweak oblique parameters~\cite{warpGHU:so5,holograph,AC}. 
On the other hand, our 4D action is suitable to see 
the whole structure of the nonlinear Higgs couplings 
among the light modes as we will see in the next section.

\subsubsection{Validity of approximations}
In our derivation of the effective action~$S_4$, 
we have taken two approximations. 
Firstly we have neglected masses of the light modes that appear 
in $S_4$ in determining their $y$-dependences. 
(See (\ref{ms_cond:gauge}) and (\ref{ms_cond:fermion}).) 
Secondly we have dropped terms suppressed by a factor of $(k\pi R)^{-1}$ 
coming from the $y$-integral of terms involving $\tl{\cA}^{\ha}_\mu$. 
To be precise, the light modes in $S_4$ 
get nonzero masses by the Higgs mechanism 
when $H^{\ha}$ have nontrivial VEV. 
The typical scale of such masses is characterized by 
the $W$ boson mass~$m_W$.\footnote{
The fermion masses are smaller than $m_W$ for $\abs{c}>1/2$. }
On the other hand, the cutoff scale of the effective theory is 
given by the KK mass scale~$\mKK$. 
In fact corrections to the expressions~(\ref{rtS4:gauge}) 
and (\ref{rtS4:fermion}) (or (\ref{rtS4:fermion2})) are estimated 
to be of or less than $\cO(\pi^2m_W^2/\mKK^2)$, 
which is $\cO(1/k\pi R)$ as we will see 
in the end of Sec.~\ref{SU3}. 
Therefore the second approximation is consistent with the first one. 
In the derivation of $S^{\rm fermi}_4$ 
we have not considered the case that $\abs{c}< 1/2$ because we do not 
obtain a simple form of the effective action by our method in such a case. 
This is related to the fact that the fermion mass becomes  
larger than $m_W$ when $\abs{c}<1/2$ and the error mentioned above 
increases.

\section{Specific models} \label{specific}
In this section we consider specific models and derive 
their effective actions by the method proposed in the previous section. 

\subsection{$\bdm{SU(3)}$ model} \label{SU3}
Here we consider the $SU(3)$ model investigated 
in Ref.~\cite{HNSS}. 
The $SU(3)$ gauge field~$A_M$ is decomposed as 
\be
 A_M = A^I_M\frac{\lmd^I}{2}, 
\ee
where $\lmd^I$ ($I=1,2,\cdots,8$) are the Gell-Mann matrices. 
As a matter field we introduce a fermion field~$\Psi$ that is 
an $SU(3)$ triplet.  
We choose $P_j$ and $\eta_j$ ($j=0,\pi$) in (\ref{orbifold_cond}) as 
\bea
 P_0 \eql P_\pi = \begin{pmatrix} -1 & & \\ & -1 & \\ & & 1 \end{pmatrix}, 
 \nonumber\\
 \eta_0 \eql \eta_\pi = +1,  
\eea
in the fundamental representation. 
Then $\bG=SU(3)$ is broken to $\bH=SU(2)\times U(1)$. 
The unbroken and the broken generators~$T^a$ and $T^{\hat{a}}$ are 
\bea
 T^a \eql \frac{\lmd^a}{2}, \;\;\;\;\; (a=1,2,3,8) \nonumber\\
 T^{\hat{a}} \eql \frac{\lmd^{\hat{a}}}{2}. \;\;\;\;\;(\hat{a}=4,5,6,7) 
\eea
The $SU(3)$-triplet~$\Psi$ is decomposed into 
\be
 \Psi =\vct{q}{Q}, 
\ee
where $q$ and $Q$ are a doublet and a singlet under 
the unbroken $SU(2)$, respectively.

\subsubsection{4D Effective action}
There appear four real scalar fields~$H^{\hat{a}}$ ($\hat{a}=4,5,6,7$) 
in low energies.  
They form an $SU(2)$-doublet and play a role of the Higgs doublet 
in the standard model. 
They do not have a potential at the classical level 
due to the 5D gauge invariance. 
So VEV of $H^{\hat{a}}$ is determined by the quantum effects, 
which is not discussed in this paper. 
Once $H^{\hat{a}}$ have a nonvanishing VEV, $SU(2)\times U(1)$ is 
broken to the $\uem$ subgroup. 
Making use of the $SU(2)\times U(1)$ symmetry of the effective action, 
such nonvanishing VEV can always be aligned to the $T^{\hat{6}}$-direction. 
Then, after the breaking of $SU(2)\times U(1)$, 
the Higgs field~$H^{\hat{6}}$ is expanded as 
\be
 H^{\hat{6}} = \fH\bthH+\tl{H},  \label{Higgs_expand:su3}
\ee
where the first and the second terms denote VEV and the fluctuation 
around it, respectively. 
In this notation, $\bthH$ becomes the VEV of the Wilson line phase, and 
\be
 \vev{\Omg_0} = \exp\brc{i\bthH\frac{\lmd^6}{2}} 
 = \begin{pmatrix} 1 & & \\ & \cw & i\sw \\ & i\sw & \cw \end{pmatrix}, 
\ee
where $\cw\equiv\cos\frac{1}{2}\bthH$ and $\sw\equiv\sin\frac{1}{2}\bthH$. 
The other scalars~$H^{\hat{a}}$ ($\hat{a}=4,5,7$) are eaten by 
the gauge bosons for $SU(2)\times U(1)/\uem$ and thus are unphysical. 
In fact, we can move to the unitary gauge in which $H^{\hat{a}}=0$ 
($\hat{a}=4,5,7$) after the breaking of $SU(2)\times U(1)$. 
Thus we focus on $H^{\hat{6}}$ among the four real scalars 
and see an explicit form of the effective action. 
The matrix~$\Omg_0$ is calculated as 
\be
 \Omg_0 = \begin{pmatrix} 1 & & \\ & \cth & i\sth \\ & i\sth & \cth 
 \end{pmatrix}+\cdots, 
 \label{Omg0_expand}
\ee
where $\cth\equiv\cos\frac{1}{2}\thH(x)$, $\sth\equiv\sin\frac{1}{2}\thH(x)$, 
and 
\be
 \thH(x) \equiv \frac{H^{\hat{6}}}{\fH}
 = \bthH+\frac{\tl{H}(x)}{\fH}.  \label{Higgs_expand2:su3}
\ee
The ellipsis in (\ref{Omg0_expand}) denotes terms involving 
$H^{\hat{a}}$ ($\hat{a}=4,5,7$). 
Then, from (\ref{def:alp_mu}), 
the ``dressed gauge fields'' and the covariant derivatives of 
the Higgs fields are read off as 
\bea
 \tl{\cA}^1_\mu+i\tl{\cA}^2_\mu \eql 
 \cth\brkt{\cA^1_\mu+i\cA^2_\mu}, \nonumber\\
 \tl{\cA}^3_\mu \eql \frac{1+\cth^2}{2}\cA^3_\mu+\frac{\sqrt{3}}{2}
 \sth^2\cA^8_\mu, \nonumber\\
 \tl{\cA}^8_\mu \eql \frac{\sqrt{3}}{2}\sth^2\cA^3_\mu
 +\frac{3\cth^2-1}{2}\cA^8_\mu, \nonumber\\
 \tl{\cD}^{(4)}_\mu\brkt{H^{\hat{4}}+iH^{\hat{5}}}
 \eql i\bar{g}_A\fH\sth\brkt{\cA^1_\mu+i\cA^2_\mu}, \nonumber\\
 \tl{\cD}^{(4)}_\mu\brkt{H^{\hat{6}}+iH^{\hat{7}}}
 \eql \der_\mu\tl{H}-i\bar{g}_A\fH\sth\cth\brkt{\cA^3_\mu-\sqrt{3}\cA^8_\mu}. 
\eea
Substituting these into (\ref{rtS4:gauge}) and (\ref{rtS4:fermion}), 
we obtain the 4D effective action. 
\bea
 S^{\rm gauge}_4 \sqa \int\dr^4x\;\left\{-\frac{1}{4}\cF^a_{\mu\nu}\cF^{a\mu\nu}
 -\frac{1}{2}\der_\mu\tl{H}\der^\mu\tl{H} 
 -\frac{\bar{g}_A^2\fH^2\sth^2}{2}\brkt{\cA^1_\mu\cA^{1\mu}+\cA^2_\mu\cA^{2\mu}}
 \right. \nonumber\\
 &&\hspace{15mm} \left. 
 -\frac{\bar{g}_A^2\fH^2\sth^2\cth^2}{2}
 \brkt{\cA^3_\mu-\sqrt{3}\cA^8_\mu}\brkt{\cA^{3\mu}-\sqrt{3}\cA^{8\mu}}
 +\cdots\right\},  \nonumber\\
 S^{\rm fermion}_4 \sqa \int\dr^4x\;i\left\{
 \bar{\psi}_L\gm^\mu\cD^{(4)}_\mu\psi_L+\bar{\chi}_R\gm^\mu\tl{\cD}^{(4)}_\mu
 \chi_R \right. \nonumber\\
 &&\hspace{15mm}\left. 
 -i\brkt{\frac{4M^2-k^2}{e^{(2M+k)\pi R}}}^{1/2}
 \brkt{\bar{\psi}_L(\Omg_0^{Qq})^\dagger\chi_R-\bar{\chi}_R\Omg_0^{Qq}\psi_L}
 +\cdots\right\}.  \label{EA:su3}
\eea
The field strengths are 
\bea
 \cF^a_{\mu\nu} \eql \der_\mu\cA^a_\nu-\der_\nu\cA^a_\mu
 +\bar{g}_A\vep_{abc}\cA^b_\mu\cA^c_\nu, \;\;\;\;\;
 (a=1,2,3) \nonumber\\ 
 \cF^8_{\mu\nu} \eql \der_\mu\cA^8_\nu-\der_\nu\cA^8_\mu, 
\eea
where $\vep_{abc}$ is the completely antisymmetric tensor of $SU(3)$, 
and the covariant derivatives and $\Omg_0^{Qq}$ are 
\bea
 \cD^{(4)}_\mu\psi_L \eql \brc{\der_\mu-i\bar{g}_A
 \sum_{a=1}^3\cA^a_\mu\frac{\sgm_a}{2}
 -i\frac{\bar{g}_A}{2\sqrt{3}}\cA^8_\mu}\psi_L, \nonumber\\
 \tl{\cD}^{(4)}_\mu\chi_R \eql \brkt{\der_\mu+\frac{i\bar{g}_A}{\sqrt{3}}
 \tl{\cA}^8_\mu}\chi_R \nonumber\\
 \eql \brc{\der_\mu
 +\frac{i\bar{g}_A}{\sqrt{3}}\brkt{\frac{\sqrt{3}}{2}\sth^2\cA^3_\mu
 +\frac{3\cth^2-1}{2}\cA^8_\mu}}\chi_R, \nonumber\\
 \Omg_0^{Qq} \eql (0,i\sth). 
\eea
Here we have assumed that $c>1/2$. 
The fields~$\psi_L$ and $\chi_R$ are a doublet and a singlet chiral spinors, 
whose components are denoted as 
\be 
 \psi_L \equiv \vct{\psi^\nu_L}{\psi^e_L}, \;\;\;\;\;
 \chi_R \equiv \chi^e_R. 
\ee

\subsubsection{Electroweak breaking phase}
After the Higgs field gets nonvanishing VEV, 
the gauge fields~($\cA^1_\mu,\cA^2_\mu,\cA^3_\mu,\cA^8_\mu$) are 
redefined to the mass eigenstates as 
\bea
 \cW_\mu \defa \frac{1}{\sqrt{2}}\brkt{\cA^1_\mu+i\cA^2_\mu}, \nonumber\\
 \vct{\cZ_\mu}{\cA^\gm_\mu} \defa \mtrx{\cos\thw}{-\sin\thw}{\sin\thw}{\cos\thw}
 \vct{\cA^3_\mu}{\cA^8_\mu}
 = \frac{1}{2}\mtrx{1}{-\sqrt{3}}{\sqrt{3}}{1}
 \vct{\cA^3_\mu}{\cA^8_\mu}.
\eea
Here $\cA^\gm_\mu$ corresponds to 
the unbroken $U(1)$ symmetry, which is identified as the photon. 
Thus the Weinberg angle~$\thw$ in this model is calculated as 
\be
 \sin\thw = \frac{\sqrt{3}}{2},  \label{thw:su3}
\ee
which is too large compared to the experimental 
value:~$\sin^2\tht^{\rm exp}_W\simeq 0.23$. 
Therefore the $SU(3)$ model cannot be a realistic model. 

The effective action after the breaking of $SU(2)\times U(1)$ becomes 
\bea
 S^{\rm gauge}_{\rm EWB} \sqa \int\dr^4x\;\left\{-\frac{1}{2}\cW_{\mu\nu}^\dagger
 \cW^{\mu\nu}-\frac{1}{4}\cF^Z_{\mu\nu}\cF^{Z\mu\nu}
 -\frac{1}{4}\cF^\gm_{\mu\nu}\cF^{\gm\mu\nu}
 -\frac{1}{2}\der_\mu\tl{H}\der^\mu\tl{H} \right. \nonumber\\
 &&\hspace{15mm}
 +i\brkt{\bar{g}\cos\thw\cF^Z_{\mu\nu}+\bar{e}\cF^\gm_{\mu\nu}}
 \cW^{\mu\dagger}\cW^\nu 
 +\frac{\bar{g}^2}{2}\brc{\abs{\cW_\mu\cW^\mu}^2
 -\brkt{\cW^\dagger_\mu\cW^\mu}^2} \nonumber\\
 &&\hspace{15mm} \left. 
 -m_W^2\cW_\mu^\dagger\cW^\mu-\frac{m_Z^2}{2}\cZ_\mu\cZ^\mu+\cdots\right\}, 
 \nonumber\\
 S^{\rm fermi}_{\rm EWB} \sqa \int\dr^4\;\left\{
 \bar{\psi}^\nu_L\gm^\mu\cD^{(4)}_\mu\psi^\nu_L
 +\bar{\psi}^e_L\gm^\mu\cD^{(4)}_\mu\psi^e_L
 +\bar{\chi}^e_R\gm^\mu\tl{\cD}^{(4)}_\mu\chi^e_R \right. \nonumber\\
 &&\hspace{15mm} \left. 
 -m_e\brkt{\bar{\psi}^e_L\chi^e_R+\bar{\chi}^e_R\psi^e_L}
 +\cdots\right\},
\eea
where the ellipses denote interaction terms with the Higgs field~$\tl{H}$. 
The other Higgs scalars~$H^{\hat{a}}$ ($\hat{a}=4,5,7$) are set to zero 
in the unitary gauge. 
The field strengths and the mass parameters are defined as 
\bea
 \cW_{\mu\nu} \defa \cD^{(4)}_\mu\cW_\nu-\cD^{(4)}_\nu\cW_\mu, \nonumber\\
 \cD^{(4)}_\mu\cW_\nu \defa \brc{\der_\mu+i\bar{g}_A\cA^3_\mu}\cW_\mu 
 = \brc{\der_\mu+i\bar{g}_A\cos\thw\cZ_\mu+i\bar{e}\cA^\gm_\mu}\cW_\mu, 
 \nonumber\\
 \cF^Z_{\mu\nu} \defa \der_\mu\cZ_\nu-\der_\nu\cZ_\mu, \;\;\;\;\;
 \cF^\gm_{\mu\nu} \equiv \der_\mu\cA^\gm_\mu-\der_\nu\cA^\gm_\nu, \nonumber\\
 m_W^2 \defa \bar{g}_A^2\fH^2\sin^2\frac{\bthH}{2}, \nonumber\\
 m_Z^2 \defa 4\bar{g}_A^2\fH^2\sw^2\cw^2
 = \bar{g}_A^2\fH^2\sin^2\bthH, \nonumber\\
 m_e \defa \brkt{\frac{4M^2-k^2}{e^{(2M+k)\pi R}}}^{1/2}\sin\frac{\bthH}{2}.  
 \label{parameters:su3}
\eea
Here $\bar{e}\equiv\bar{g}_A\sin\thw$ is the $\uem$ gauge coupling. 

The expressions of the mass parameters in (\ref{parameters:su3}) agree with 
those derived by the KK analysis. 
The corrections to them are estimated as $\cO(\pi^2m_f^2/\mKK^2)$ 
for $m_f^2$ ($f=W,Z,e$). 
(See Eq.(5.5) in Ref.~\cite{HNSS}.)
For the $W$ boson, for example, this ratio becomes  
\be
 \frac{\pi^2m_W^2}{\mKK^2} \simeq \frac{1}{k\pi R}\sin^2\frac{\bthH}{2}, 
\ee 
which means a few percents error for $e^{k\pi R}=\cO(10^{15})$ 
and $\bthH=\cO(1)$. 

Note that the masses are not proportional to the ``Higgs VEV''~$\bthH$. 
This nonlinearity comes from the nonlinear structure 
of the Higgs couplings in (\ref{EA:su3}). 
Furthermore the $\bthH$-dependence of $m_W$ and $m_Z$ are different. 
So the $\rho$ parameter depends on $\bthH$ and deviates from one 
for general values of $\bthH$. 
\be
 \rho \equiv \frac{m_W^2}{m_Z^2\cos^2\thw} = \frac{1}{\cos^2\frac{1}{2}\bthH}. 
\ee
This is another problem of the $SU(3)$ model. 

The interaction terms will be discussed in more realistic model 
considered in the next subsection.

\subsection{$\bdm{SO(5)\times\ubl}$ model}
Here we consider the $SO(5)\times\ubl$ model 
that is analyzed in our previous papers~\cite{SH,HS1}.\footnote{
This type of model is first studied in Ref.~\cite{warpGHU:so5}. }
We have the $SO(5)$ gauge field~$A_M$ and the $\ubl$ gauge field~$B_M$. 
The former is decomposed as 
\be
 A_M=A^I_MT^I, 
\ee
where $T^I$ ($I=1,2,\cdots,10$) are the generators of $SO(5)$. 
The spinorial representation of $T^I$ is tabulated 
in (\ref{SO_generator}) in Appendix~\ref{notation}. 
As a matter field we introduce a fermion field~$\Psi$ 
in the spinorial representation of $SO(5)$ (\ie, $\bdm{4}$ of SO(5)). 
We choose $P_j$ and $\eta_j$ ($j=0,\pi$) in (\ref{orbifold_cond}) as 
\bea
 P_0 \eql P_\pi = \mtrx{1_2}{}{}{-1_2}, \nonumber\\
 \eta_0 \eql \eta_\pi = -1. 
\eea
Then $\bG=SO(5)\times\ubl$ is broken to 
$\bH=SO(4)\times\ubl\sim\suL\times\suR\times\ubl$. 
The unbroken generators are the $\suL\times\suR$ 
generators~$(T^{\aL},T^{\aR})$ ($\aL,\aR=1,2,3$) 
and the $\ubl$ generator~$\cQ_{\rm B-L}$, 
while the broken ones are $T^{\hat{a}}$ ($\hat{a}=1,2,3,4$). 
The fermion~$\Psi$ is decomposed as 
\be
 \Psi = \vct{q}{Q}, 
\ee
where $q$ and $Q$ belong to $(\bdm{2},\bdm{1})$ and 
$(\bdm{1},\bdm{2})$ of $\suL\times\suR$, respectively. 
The covariant derivative of $\Psi$ is defined as 
\be
 \cD_M\Psi \equiv \brkt{\der_M-\frac{1}{4}\omg_M^{\;\;AB}\Gm_{AB}
 -ig_A A^I_MT^I-i\frac{g_B}{2}B_M\cQ_{\rm B-L}}\Psi, 
\ee
where $I=(\aL,\aR,\hat{a})$. 

In contrast to the previous $SU(3)$ model, 
the unbroken gauge symmetry~$\suL\times\suR\times\ubl$ is too large 
to be identified as the electroweak symmetry. 
In Ref.~\cite{SH}, we add an additional dynamics on the UV brane 
in order to construct a realistic model. 
It breaks $\suR\times\ubl$ to a subgroup~$\uy$ spontaneously 
at a relatively high energy scale~$M_{\rm R}$.  
Then the following mass temrs are induced on the UV brane 
below the scale~$M_{\rm R}$. 
\be
 \cL_{\rm mass} = -\brc{M_1^2\brkt{A^{1_{\rm R}}_\mu A^{1_{\rm R}\mu}
 +A^{2_{\rm R}}_\mu A^{2_{\rm R}\mu}}
 +M_2^2A^{'3_{\rm R}}_\mu A^{'3_{\rm R}\mu}}\dlt(y), 
 \label{UV_mass}
\ee
where $M_1, M_2 = \cO(M_{\rm R})$, and 
\be
 \vct{A^{'3_{\rm R}}_\mu}{A^Y_\mu} \equiv \mtrx{\cph}{-\sph}{\sph}{\cph}
 \vct{A^{3_{\rm R}}_\mu}{B_\mu}, \nonumber
\ee
\be
 \cph \equiv \frac{g_A}{\sqrt{g_A^2+g_B^2}}, \;\;\;\;\;
 \sph \equiv \frac{g_B}{\sqrt{g_A^2+g_B^2}}. 
\ee
The constants~$g_A$ and $g_B$ are the gauge couplings for $SO(5)$ and 
$\ubl$, respectively. 
The mass terms~(\ref{UV_mass}) effectively change the boundary conditions 
for $(A^{1_{\rm R}}_\mu, A^{2_{\rm R}}_\mu, A^{'3_{\rm R}}_\mu)$ 
at the UV brane from the Neumann to the Dirichlet conditions. 
This does not cause an essential change 
of the derivation of the effective theory in Sec.~\ref{derivation_S4}. 
Recall that the solutions~(\ref{gen_sol:gauge}) and (\ref{gen_sol:fermion}) 
are determined only by the boundary conditions at the IR brane. 
The only effect of $\cL_{\rm mass}$ in (\ref{UV_mass}) is 
to force the field values of 
$(A^{1_{\rm R}}_\mu, A^{2_{\rm R}}_\mu, A^{'3_{\rm R}}_\mu)$ 
at the UV brane down to zero. 
Namely, 
\bea
 \bar{g}_A\brkt{\cA^{\aL}_\mu T^{\aL}+\cA^{\aR}_\mu T^{\aR}}
 +\frac{\bar{g}_B}{2}\cQ_{\rm B-L}\cB_\mu 
 \eql \bar{g}_A\brkt{\cA^{\aL}_\mu T^{\aL}+\sph\cA^Y_\mu T^{3_{\rm R}}}
 +\frac{\bar{g}_B}{2}\cQ_{\rm B-L}(\cph\cA^Y_\mu) \nonumber\\
 \eql \bar{g}\cA^{\aL}_\mu T^{\aL}+\bar{g}'\cQ_Y\cA^Y_\mu, 
\eea
where $\bar{g}\equiv\bar{g}_A$, $\bar{g}'\equiv\sph\bar{g}_A=\cph\bar{g}_A$ 
are the 4D gauge couplings for $\suL$ and $\uy$ respectively,  
and $\cQ_Y\equiv T^{3_{\rm R}}+\cQ_{\rm B-L}/2$ 
is the charge of $\uy$. 
Thus we can obtain the 4D effective action of this model 
by setting the 4D gauge fields~$(\cA^{1_{\rm R}}_\mu,\cA^{2_{\rm R}}_\mu,
\cA^{'3_{\rm R}}_\mu)$ to zero at the last step of the procedure.

\subsubsection{4D Effective action}
There appear four real scalars~$H^{\hat{a}}$ ($\hat{a}=1,2,3,4$) 
in low energies. 
They form an $\suL$-doublet and play a role of 
the Higgs doublet in the standard model. 
Once $H^{\hat{a}}$ have nonvanishing VEV, 
$\suL\times\uy$ is broken to the electromagnetic gauge group~$\uem$. 
Making use of the $\suL\times\uy$ symmetry of the effective action, 
such VEV can always be aligned 
along the $T^{\hat{4}}$-direction. 
Then the Higgs field~$H^{\hat{4}}$ is expanded as 
\be
 H^{\hat{4}} = \sqrt{2}\fH\bthH+\tl{H}, 
\ee
where the first and the second terms denote VEV and the fluctuation field, 
respectively. 
\ignore{
In this notation, $\bthH$ becomes the VEV of the Wilson line phase, and 
\be
 \vev{\Omg_0} = \exp\brc{\frac{i}{2}\bthH (2\sqrt{2}T^{\hat{4}})}
 = \mtrx{\cw}{i\sw}{i\sw}{\cw}\otimes 1_2, 
\ee
where $\cw\equiv\cos\frac{1}{2}\bthH$ 
and $\sw\equiv\sin\frac{1}{2}\bthH$. }
The other scalars~$H^{\hat{a}}$ ($\hat{a}=1,2,3$) are eaten by 
the gauge bosons and thus unphysical. 
In fact, we can move to the unitary gauge in which $H^{\hat{a}}=0$ 
($\hat{a}=1,2,3$) after the breaking of $\suL\times\uy$. 
Thus we focus on $H^{\hat{4}}$ among the four real scalars 
and see an explicit form of the effective action. 
The matrix~$\Omg_0$ is calculated as 
\be
 \Omg_0 = \mtrx{\cth}{i\sth}{i\sth}{\cth}\otimes 1_2+\cdots, 
 \label{Omg0}
\ee
where $\cth\equiv\cos\frac{1}{2}\thH(x)$, $\sth\equiv\sin\frac{1}{2}\thH(x)$, 
and 
\be
 \thH(x) \equiv \frac{H^{\hat{4}}(x)}{\sqrt{2}\fH}
 = \bthH+\frac{\tl{H}(x)}{\sqrt{2}\fH}. 
\ee
The ellipses in (\ref{Omg0}) and in the following expressions 
denote terms involving $H^{\hat{a}}$ ($\hat{a}=1,2,3$). 
\ignore{
Then the gauged Maurer Cartan 1 form~$\alp_\mu$ is calculated as 
\bea
 \alp_\mu \eql \Omg_0\brkt{\bar{g}_A\cA^{\aL}_\mu T^{\aL}
 +\bar{g}_A\cA^{\aR}_\mu T^{\aR}+\frac{\bar{g}_B}{2}
 q_{\rm B-L}\cB_\mu}\Omg_0^{-1}-i\der_\mu\Omg_0\Omg_0^{-1}  \nonumber\\
 \eql \frac{1}{2}
 \mtrx{\bar{g}_A\brkt{\cth^2\cA^{\aL}_\mu+\sth^2\cA^{\aR}_\mu}\sgm_a
 +\bar{g}_Bq_{\rm B-L}\cB_\mu}
 {i\der_\mu\tht-i\sth\cth\bar{g}_A
 \brkt{\cA^{\aL}_\mu-\cA^{\aR}_\mu}\sgm_a}
 {i\der_\mu\tht+i\sth\cth\bar{g}_A
 \brkt{\cA^{\aL}_\mu-\cA^{\aR}_\mu}\sgm_a}
 {\bar{g}_A\brkt{\sth^2\cA^{\aL}_\mu+\cth^2\cA^{\aR}_\mu}\sgm_a
 +\bar{g}_Bq_{\rm B-L}\cB_\mu} \nonumber\\
 &&+\cdots. 
\eea}
From the gauged Maurer Cartan 1 form~$\alp_\mu$ defined 
in (\ref{def:alp_mu}), we can read off the ``dressed gauge fields'' and 
the covariant derivatives of the Higgs fields as 
\bea
 \tl{\cA}^{\aL}_\mu \eql \cth^2\cA^{\aL}_\mu+\sth^2\cA^{\aR}_\mu
 +\cdots,  \nonumber\\
 \tl{\cA}^{\aR}_\mu \eql \sth^2\cA^{\aL}_\mu+\cth^2\cA^{\aR}_\mu
 +\cdots,  \nonumber\\
 \tl{\cB}_\mu \eql \cB_\mu, \nonumber\\
 \tl{\cD}^{(4)}_\mu H^{\hat{a}} \eql -\sqrt{2}\bar{g}_A\fH\sth\cth
 \brkt{\cA^{\aL}_\mu-\cA^{\aR}_\mu}+\cdots, \;\;\;\;\; 
 (\aL=\aR=\hat{a}=1,2,3) \nonumber\\
 \tl{\cD}^{(4)}_\mu H^{\hat{4}} \eql \sqrt{2}\fH\der_\mu\thH
 = \der_\mu\tl{H}+\cdots. 
\eea
Substituting these into (\ref{rtS4:gauge}) and (\ref{rtS4:fermion}), 
the following expressions are obtained. 
\bea
 S^{\rm gauge}_4 \sqa \int\dr^4x\;\left\{
 -\frac{1}{4}\brkt{\cF^{\aL}_{\mu\nu}\cF^{\aL\mu\nu}
 +\cF^{\aR}_{\mu\nu}\cF^{\aR\mu\nu}+\cF^B_{\mu\nu}\cF^{B\mu\nu}}
 \right. \nonumber\\
 &&\hspace{15mm}\left. 
 -\frac{1}{2}\der_\mu\tl{H}\der^\mu\tl{H}
 -\bar{g}_A^2\fH^2\sth^2\cth^2\brkt{\cA^{\aL}_\mu-\cA^{\aR}_\mu}
 \brkt{\cA^{\aL\mu}-\cA^{\aR\mu}}\right\}+\cdots, \nonumber\\
 S^{\rm fermi}_4 \sqa \int\dr^4x\;i\left\{
 \bar{\psi}_L\gm^\mu\brkt{\der_\mu-i\bar{g}_A\cA^{\aL}_\mu
 \frac{\sgm_a}{2}
 -\frac{\bar{g}_B}{2}\cB_\mu\cQ_{\rm B-L}}\psi_L \right. \nonumber\\
 &&\hspace{15mm}
 +\bar{\chi}_R\gm^\mu\brkt{\der_\mu-i\bar{g}_A\brkt{
 \sth^2\cA^{\aL}_\mu+\cth^2\cA^{\aR}_\mu}\frac{\sgm_a}{2}
 -\frac{\bar{g}_B}{2}\cB_\mu\cQ_{\rm B-L}}\chi_R  \nonumber\\
 &&\hspace{15mm}\left. 
 -\brkt{\frac{4M^2-k^2}{e^{(2M+k)\pi R}}}^{1/2}\sth
 \brkt{\bar{\psi}_L\chi_R+\bar{\chi}_R\psi_L}\right\}+\cdots,
\eea
where $\cF^{\aL,\aR}_{\mu\nu}$ and $\cF^B_{\mu\nu}$ are 
the field strengths of $\cA^{\aL,\aR}_\mu$ and $\cB_\mu$, respectively. 
We have assumed that $c>1/2$. 
By setting $(\cA^{1_{\rm R}}_\mu,\cA^{2_{\rm R}}_\mu,
\cA^{'3_{\rm R}}_\mu)$ to zero in the above expressions, 
we obtain the 4D effective action. 
\bea
 S^{\rm gauge}_4 \sqa \int\dr^4x\;\left\{
 -\frac{1}{4}\brkt{\cF^{\aL}_{\mu\nu}\cF^{\aL\mu\nu}
 +\cF^Y_{\mu\nu}\cF^{Y\mu\nu}} 
 -\frac{1}{2}\der_\mu\tl{H}\der^\mu\tl{H} \right. \nonumber\\
 &&\hspace{15mm}\left.
 -\frac{\bar{g}^2\fH^2\sin^2\thH}{4}
 \brc{\cA^{1_{\rm L}}_\mu\cA^{1_{\rm L}\mu}
 +\cA^{2_{\rm L}}_\mu\cA^{2_{\rm L}\mu}
 +(\cA^{3_{\rm L}}_\mu-\sph\cA^Y_\mu)(\cA^{3_{\rm L}\mu}-\sph\cA^{Y\mu})}
 \right\}, \nonumber\\
 &&+\cdots, \nonumber\\
 S^{\rm fermi}_4 \sqa \int\dr^4x\;i\brc{
 \bar{\psi}_L\gm^\mu\cD^{(4)}_\mu\psi_L 
 +\bar{\chi}_R\gm^\mu\tl{\cD}^{(4)}_\mu\chi_R  
 -\brkt{\frac{4M^2-k^2}{e^{(2M+k)\pi R}}}^{1/2}\sth
 \brkt{\bar{\psi}_L\chi_R+\bar{\chi}_R\psi_L}}, \nonumber\\
 &&+\cdots,  \label{S4:so5}
\eea
where $\cF^Y_{\mu\nu}$ is the field strength of $\cA^Y_\mu$, and 
\bea
 \cD^{(4)}_\mu\psi_L \eql \brc{\der_\mu
 -i\bar{g}\cA^{\aL}_\mu\frac{\sgm_{\aL}}{2}
 -i\bar{g}'\cA^Y_\mu\frac{\cQ_{\rm B-L}}{2}}\psi_L, \nonumber\\
 \tl{\cD}^{(4)}_\mu\chi_R \eql \brc{\der_\mu
 -i\bar{g}\sth^2\cA^{\aL}_\mu\frac{\sgm_{\aL}}{2}
 -i\bar{g}'\cA^Y_\mu\brkt{\cth^2\frac{\sgm_3}{2}
 +\frac{q_{\rm B-L}}{2}}}\chi_R. 
 \label{def_cD:so5}
\eea
Note that $\cQ_Y=\cQ_{\rm B-L}/2$ on $\psi_L$ since $T^{3_{\rm R}}=0$.

\subsubsection{Electroweak breaking phase}
After the Higgs field gets nonvanishing VEV, 
the mass eigenstates become 
\bea
 \cW_\mu \defa \frac{1}{\sqrt{2}}\brkt{
 \cA^{1_{\rm L}}_\mu+i\cA^{2_{\rm L}}_\mu}, \nonumber\\
 \vct{\cZ_\mu}{\cA^\gm_\mu} \defa \mtrx{\cos\thw}{-\sin\thw}{\sin\thw}{\cos\thw}
 \vct{\cA^{3_{\rm L}}_\mu}{\cA^Y_\mu} 
 = \frac{1}{\sqrt{1+\sph^2}}
 \mtrx{1}{-\sph}{\sph}{1}\vct{\cA^{3_{\rm L}}_\mu}{\cA^Y_\mu}. 
\eea
Thus the Weinberg angle~$\thw$ is read off as 
\be
 \tan\thw = \sph. 
\ee
Then the effective action in the electroweak breaking phase becomes
\bea
 S^{\rm gauge}_{\rm EWB} \sqa \int\dr^4x\;\left\{
 -\frac{1}{2}\cW_{\mu\nu}^\dagger\cW^{\mu\nu}
 -\frac{1}{4}\cF^Z_{\mu\nu}\cF^{Z\mu\nu}
 -\frac{1}{4}\cF^\gm_{\mu\nu}\cF^{\gm\mu\nu}
 -\frac{1}{2}\der_\mu\tl{H}\der^\mu\tl{H} \right. \nonumber\\
 &&\hspace{15mm}
 +i\brkt{\bar{g}\cos\thw\cF^Z_{\mu\nu}+\bar{e}\cF^\gm_{\mu\nu}}
 \cW^{\mu\dagger}\cW^\nu 
 +\frac{\bar{g}^2}{2}\brc{\abs{\cW_\mu\cW^\mu}^2
 -\brkt{\cW^\dagger_\mu\cW^\mu}^2} \nonumber\\
 &&\hspace{15mm}\left. 
 -\frac{\bar{g}^2\fH^2}{2}\sin^2\thH\cW^\dagger_\mu\cW^\mu
 -\frac{\bar{g}^2\fH^2}{4\cos^2\thw}\sin^2\thH\cZ_\mu\cZ^\mu\right\}, 
 \nonumber\\
 S^{\rm fermi}_{\rm EWB} \sqa \int\dr^4x\;i\left\{
 \bar{\psi}_L\gm^\mu\cD^{(4)}_\mu\psi_L
 +\bar{\chi}_R\gm^\mu\tl{\cD}^{(4)}_\mu\chi_R  \right. \nonumber\\
 &&\hspace{15mm}\left. 
 -\brkt{\frac{4M^2-k^2}{e^{(2M+k)\pi R}}}^{1/2}\sin\frac{\thH}{2}
 \brkt{\bar{\psi}_L\chi_R+\bar{\chi}_R\psi_L}\right\}, 
 \label{EA:so5}
\eea
where $\bar{e}\equiv\bar{g}\sin\thw$ is the $\uem$ gauge coupling. 
\ignore{
The field strengths are defined as  
\bea
 \cW_{\mu\nu} \defa \cD^{(4)}_\mu\cW_\nu-\cD^{(4)}_\nu\cW_\mu, 
 \nonumber\\
 \cD^{(4)}_\mu\cW_\nu \defa \brc{\der_\mu
 +i\bar{g}\cA^{3_{\rm L}}_\mu}\cW_\mu 
 = \brc{\der_\mu+i\bar{g}\cos\thw\cZ_\mu+i\bar{e}\cA^\gm_\mu}\cW_\mu, 
 \nonumber\\
 \cF^Z_{\mu\nu} \defa \der_\mu\cZ_\nu-\der_\nu\cZ_\mu, \;\;\;\;\;
 \cF^\gm_{\mu\nu} \equiv \der_\mu\cA^\gm_\nu-\der_\nu\cA^\gm_\mu. 
\eea }
We have taken the unitary gauge in which $H^{\hat{a}}=0$ ($\hat{a}=1,2,3$).  
The definitions of the field strengths are the same as 
in (\ref{parameters:su3}) but now 
\be
  \cD^{(4)}_\mu\cW_\nu \equiv \brc{\der_\mu
 +i\bar{g}\cA^{3_{\rm L}}_\mu}\cW_\mu 
 = \brc{\der_\mu+i\bar{g}\cos\thw\cZ_\mu+i\bar{e}\cA^\gm_\mu}\cW_\mu. 
\ee
From the couplings to the Higgs field, 
the mass parameters are read off as 
\bea
 m_W^2 \defa \frac{\bar{g}^2\fH^2\sin^2\bthH}{2}, \nonumber\\
 m_Z^2 \defa \frac{\bar{g}^2\fH^2\sin^2\bthH}{2\cos^2\thw} 
 = \frac{m_W^2}{\cos^2\thw}, 
 \nonumber\\
 m_e \defa \brkt{\frac{4M^2-k^2}{e^{(2M+k)\pi R}}}^{1/2}\sin\frac{\bthH}{2}. 
 \label{parameters:so5}
\eea
These expressions agree with those derived 
by the KK analysis~\cite{SH}. 
In contrast to the $SU(3)$ model, the $W$ and the $Z$ boson masses have 
the same $\bthH$-dependence so that the $\rho$ parameter is now 
independent of $\bthH$ and equals one, which is consistent with 
the experiments. 
This can be understood as a result of 
the custodial symmetry. 
Note that the Higgs fields~$H^{\hat{a}}$ ($\hat{a}=1,2,3,4$) 
form a doublet not only for $\suL$ but also for $\suR$. 
Thus the Higgs sector of 
$S^{\rm gauge}_4$ in (\ref{S4:so5}) has a global symmetry 
$\suL\times\suR$ if we set $\sph=0$ (or $g_B=0$). 
After $H^{\hat{4}}$ gets VEV, this global symmetry is broken to 
its diagonal subgroup~$SU(2)_{\rm D}$. 
This custodial symmetry ensures $\rho=1$.

\subsubsection{Interaction terms} \label{int_terms}
Now we discuss the interaction terms. 
Firstly we can immediately see from (\ref{EA:so5}) 
that self-interactions of the gauge fields, 
such as the $WWZ$, $WWWW$ and $WWZZ$ couplings,  
are the same as those of the standard model. 
This is consistent with the results obtained 
by the KK analysis~\cite{SH,HS1}. 

\begin{figure}[t,b]
\centering  \leavevmode
\includegraphics[width=80mm]{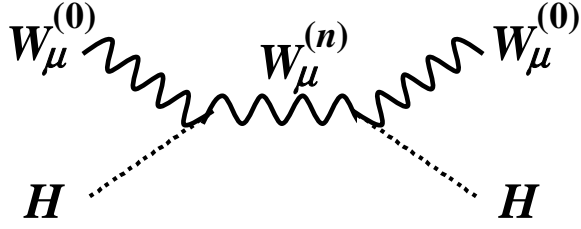}
\caption{An additional contribution to the $WWHH$ coupling. 
$W^{(n)}_\mu$ denotes the $n$-th KK excitation mode 
of the $W$ boson. 
}
\label{ad_diagram}
\end{figure}

Next we investigate the couplings among the gauge 
and the Higgs fields, \ie, 
the third line of $S^{\rm gauge}_{\rm EWB}$ in (\ref{EA:so5}). 
By expanding the corresponding terms 
in terms of the fluctuation~$\tl{H}$ around VEV, we obtain 
\bea
 \cL^{\rm int}_4 \eql -\frac{\bar{g}^2\fH^2}{2}
 \sin^2\brkt{\bthH+\frac{\tl{H}}{\sqrt{2}\fH}}
 \cW^\dagger_\mu\cW^\mu+\cdots \nonumber\\
 \eql -\frac{\bar{g}^2\fH^2}{2}
 \brc{\sin^2\bthH+2\sin\bthH\cos\bthH\cdot 
 \frac{\tl{H}}{\sqrt{2}\fH}+\cos 2\bthH\cdot\frac{\tl{H}^2}{2\fH^2}
 +\cO(\tl{H}^3)}\cW^\dagger_\mu\cW^\mu+\cdots  \nonumber\\
 \eql -m_W^2\cW^\dagger_\mu\cW^\mu
 -\lmd_{WWH}\cW^\dagger_\mu\cW^\mu\tl{H}
 -\frac{\lmd^2_{WWHH}}{4}\cW^\dagger_\mu\cW^\mu\tl{H}^2
 +\cdots.  \label{int:H-W}
\eea
where 
\bea
 \lmd_{WWH} \eql \bar{g}m_W\cos\bthH, \nonumber\\
 \lmd^2_{WWHH} \eql \bar{g}^2\cos 2\bthH.  \label{lmds}
\eea
The $WWH$ coupling~$\lmd_{WWH}$ is consistent with the result 
obtained in the KK analysis~\cite{SH,HS1}. 
It is suppressed by a factor~$\cos\bthH$ compared to the counterpart 
in the standard model. 
This suppression factor can be easily understood from a nonlinear structure 
in the Higgs sector of the effective action. 
Eq.(\ref{lmds}) shows that the $WWHH$ coupling is suppressed 
by a factor~$\cos 2\bthH$ compared to the standard model. 
This seems to contradict the result 
obtained in Ref.~\cite{HS1} where the suppression factor is estimated as 
$(1-\frac{2}{3}\sin^2\bthH)$. 
However we have to notice that the couplings calculated in Ref.~\cite{HS1} 
are the {\it bare} couplings~$\lmd_{WWHH}^{2\;\rm bare}$. 
In energies below the compactification scale~$\mKK$, 
the massive KK modes are integrated out and induce additional 
contributions to some couplings among the light modes. 
In fact a diagram depicted in Fig.~\ref{ad_diagram} also contributes 
to the $WWHH$ couplings in the low-energy effective theory. 
As shown in the appendix~\ref{cal_WWHH}, this additional contribution 
to $\lmd^{2\;\rm bare}_{WWHH}$ is estimated as 
\be
 \dlt\lmd^2_{WWHH} \simeq -\frac{4}{3}\bar{g}^2\sin^2\bthH. 
\ee
Thus the effective $WWHH$ coupling~$\lmd^{\rm eff}_{WWHH}$ becomes 
\bea
 \lmd^{2\;\rm eff}_{WWHH} \eql \lmd^{2\;\rm bare}_{WWHH}
 +\dlt\lmd^2_{WWHH} \nonumber\\
 \sqa \bar{g}^2\brkt{1-2\sin^2\bthH} = \bar{g}^2\cos 2\bthH, 
\eea
which is consistent with (\ref{lmds}). 
A situation is similar for the $ZZH$ and the $ZZHH$ couplings. 

Finally we discuss the fermion sector. 
The effective action~(\ref{EA:so5}) reproduces the results of 
Ref.~\cite{SH,HS1} also for the gauge couplings of the fermions. 
From (\ref{def_cD:so5}), the covariant derivative 
of $\psi_L$ is written in the electroweak breaking phase as 
\bea
 \cD^{(4)}_\mu\psi_L \eql \left\{\der_\mu-\frac{i\bar{g}}{\sqrt{2}}
 \mtrx{}{\cW^\dagger_\mu}{\cW_\mu}{} \right. \nonumber\\
 &&\hspace{5mm} \left.
 -\frac{i\bar{g}}{\cos\thw}\brkt{\frac{\sgm_3}{2}
 -\sin^2\thw\cQ_{\rm EM}}\cZ_\mu
 -i\bar{e}\cA^\gm_\mu\cQ_{\rm EM}\right\}
 \vct{\psi^\nu_L}{\psi^e_L}, 
\eea
where $\cQ_{\rm EM}\equiv T^{3_{\rm L}}+\cQ_Y 
=T^{3_{\rm L}}+T^{3_{\rm R}}+\cQ_{\rm B-L}/2$ is 
the $\uem$ charge operator. 
This agrees with that of the standard model, 
and is consistent with the experiments. 
On the other hand, the gauge couplings for $\chi_R$ deviate 
from the standard model. 
The covariant derivative of $\chi_R$ becomes  
\bea
 \tl{\cD}^{(4)}_\mu\chi_R \eql \left\{\der_\mu-i\frac{\bar{g}}{\sqrt{2}}
 \sth^2\mtrx{}{\cW^\dagger_\mu}{\cW_\mu}{} \right. \nonumber\\
 &&\hspace{5mm} \left. 
 -\frac{i\bar{g}}{\cos\thw}\brkt{\sth^2\frac{\sgm_3}{2}-\sin^2\thw\cQ_{\rm EM}}
 \cZ_\mu-i\bar{e}\cA^\gm_\mu\cQ_{\rm EM} \right\}
 \vct{\chi^\nu_R}{\chi^e_R}. 
\eea
The couplings to the $W$ and the $Z$ bosons substantially deviate from 
the standard model values for $\bthH=\cO(1)$. 
Especially the right-handed fermion~$\chi_R$ couples to the $W$ boson. 
The situation is similar also in the case of $c<-1/2$. 
The gauge couplings for the left-handed fermion~$\chi_L$ deviate 
from the standard model values in that case. 
From the viewpoint of our effective action, 
these deviations stem from the fact that the gauge fields that 
couple to the modes localized near the IR brane are dressed 
by the Higgs field. 
Therefore the gauge couplings of the fermions inevitably 
deviate from the standard model values unless they are localized 
near the UV brane. 
This problem can be solved by introducing some additional chiral fermions 
on the orbifold boundaries and mixing them with the bulk fermion~$\Psi$. 
This possibility will be investigated thoroughly in Ref.~\cite{HS2}. 

The suppression of the Yukawa couplings, 
which is discussed in the $SU(3)$ model in Ref.~\cite{HNSS}, 
can also be explained by the nonlinear structure of 
the Higgs sector in the effective action~(\ref{EA:su3}).

\section{Summary} \label{summary}
We have derived the 4D effective actions
of the 5D gauge-Higgs unification models in the warped spacetime. 
The derivation is simple and straightforward. 
To extract light modes that appear in the effective theory 
from the 5D fields, we imposed the ``massless conditions'' 
(\ref{ms_cond:gauge}) and (\ref{ms_cond:fermion}). 
Although these modes can obtain nonzero masses 
after the Higgs fields have nontrivial VEV, 
our prescription is a good approximation 
because the their masses~$m_f$ are much lighter 
than the KK excitation scale~$\mKK$ in the warped spacetime. 
In the determination of the mode functions of the light modes 
in the KK analysis, 
the effects of $m_f$ are suppressed by $m_f^2/\mKK^2$ 
and subdominant. 
Imposing the ``massless conditions'' corresponds 
to just neglecting such subdominant effects. 
The nonzero masses~$m_f$ are safely reproduced with a good approximation 
through the couplings to the Higgs fields. 
The ``massless conditions'' greatly simplifies the derivation 
of the effective theory because they determine the mode functions 
of the light modes without using the detailed information 
on the mass spectrum. 
Notice that only the boundary conditions at the IR brane are necessary 
to determine the mode functions. 
The role of those at the UV brane is 
to set the UV-boundary values of some 5D fields to zero 
and select 4D fields that are dynamical in the effective theory. 


The nonlinear structure of the Higgs couplings can be 
explicitly seen by deriving the 4D action 
in the gauge where $A_y=0$. 
The resultant 4D action is obtained by calculating 
the gauged Maurer Cartan 1-form~$\alp_\mu$. 
All the coupling constants can be read off from 
the formulae~(\ref{rtS4:gauge}), (\ref{rtS4:fermion}) 
(or (\ref{rtS4:fermion2})).  
The self-interactions among the gauge fields are the same as 
those in the standard model. 
The Higgs couplings to the $W$ and the $Z$ bosons are suppressed by 
factors that depend on the Wilson line phase~$\bthH$ 
from the standard model. 
We can directly read off those suppression factors 
which we have obtained by somewhat complicated calculations 
in Ref.~\cite{HNSS,SH,HS1}. 
The problematic deviations of the fermionic gauge couplings  
are also manifest in our effective  action. 
These are the advantage of our approach.  

\ignore{
In the KK analysis 
such nonlinear structure of the Higgs couplings 
is not manifest to see. 
The 4D Higgs scalar mode comes from the fluctuation 
of the fifth component of the gauge potential~$A_y$ 
around the background field configuration. 
Therefore we cannot see the nonlinear structure of the Higgs couplings 
because the original 5D action has terms involving $A_y$ 
only up to quartic.  
To see the nonlinear structure, we gauged $A_y$ away at the first step 
and pushed its degrees of freedom 
into the Wilson line phase~$\tht^{\ha}(x,\ep)$. }

The gauge group~$\bG$ is broken by the boundary conditions 
at both boundaries. 
From the viewpoint of the effective theory, 
the symmetries broken at the IR brane are realized nonlinearly 
while those at the UV brane are explicitly broken. 
In fact the effective action has the same structure as those 
of 4D models in which the Higgs fields are provided as 
the pseudo Nambu-Goldstone (PNG) bosons. 
These properties are consistent 
with the holographic interpretation~\cite{LPR,holograph,AC,Panico}. 

The Higgs field~$\tl{H}$ in this paper 
and the scalar zero-mode~$H^{(0)}$ in Ref.~\cite{SH} 
represent the same degree of freedom. 
However they have different forms of the interactions. 
The former has interaction terms at any order 
while the latter does not. 
In fact the latter has only up to quartic couplings 
since it comes from the fluctuation mode of 
the fifth component of the gauge potential~$A_y$. 
Furthermore they have different couplings to the $W$ and the $Z$ bosons 
as we have seen in Sec.~\ref{int_terms}. 
The latter has nonvanishing $W^{(0)}W^{(n)}H^{(0)}$ couplings ($n\neq 0$) 
that induce the correction to the $WWHH$ coupling. 
This means that we cannot simply drop the KK excitation modes 
to obtain the effective action. 
On the other hand, $\tl{H}$ has the correct value of the $WWHH$ coupling 
in the effective action. 
Therefore $\tl{H}$ is regarded as a field obtained by the field redefinition 
of $H^{(0)}$: 
\be
 \tl{H} = H^{(0)}+\cO(H^{(0)2}), 
\ee
so that the $W^{(0)}W^{(n)}\tl{H}$ coupling vanishes. 

Finally we comment on the limits of validity for our approach. 
First we should emphasize that our method is valid only 
in the warped spacetime. 
In the flat limit ($k\pi R\to 0$), the KK mass scale~$\mKK$ 
becomes closer to the electroweak scale~$m_W$ unless $\bthH\ll 1$. 
(See Fig.~1 of Ref.~\cite{HNSS} or Fig.~1 of Ref.~\cite{HS1}.) 
Therefore we cannot neglect the KK modes like we did in this paper. 
They provide non-negligible effects to the low energy physics 
when they are integrated out. 
It is known that $\bthH$ must be tiny to avoid too light Higgs boson 
in the flat spacetime. 
When $\bthH\ll 1$, our method is applicable even in the flat case, 
but the nonlinear structure of the Higgs sector almost disappears 
and the effective theory is reduced to the ordinary standard model. 
When the warp factor is large enough (\ie, $k\pi R\gg 1$), 
the effects of the KK modes are negligible 
and our method is safely applied. 
In fact the corrections of the mass formulae in (\ref{parameters:su3}) 
and (\ref{parameters:so5}) are of order or less than $\cO(1/k\pi R)$. 
\ignore{
The $WWZ$ couplings are close to the standard model values 
with much better accuracy, \ie, less than $\cO(10^{-4})$ 
for $k\pi R\simeq 35$. 
(See Table~III in Ref.~\cite{HS1}.)
This indicates the existence of some mechanism that protects 
the self couplings of the gauge bosons to the standard model values. }

\vskip 0.5cm

\leftline{\bf Acknowledgments} \nopagebreak
The author would like to thank Y. Hosotani for useful discussions and comments. 
This work was supported by JSPS fellowship No.\ 0509241.

\appendix

\section{Notations} \label{notation}
The metric convention of the 4D Minkowski space is taken as
\be
 \eta_{\mu\nu} = \diag(-1,1,1,1), 
\ee
and the Clifford algebra is given by 
\be
 \brc{\Gm^A,\Gm^B} = 2\eta^{AB}. 
\ee
An explicit representation of the $\gm$-matrices is given by  
\be
 \Gm^0 = \gm^0 = \mtrx{}{i1_2}{i1_2}{}, \;\;\;\;\;
 \Gm^j = \gm^j = \mtrx{}{i\sgm_j}{-i\sgm_j}{}, \;\;\;\;\;
 \Gm^4 = \gm_5 = \mtrx{1_2}{}{}{-1_2},  \label{Gm_rep}
\ee
where $j=1,2,3$ and $\sgm_j$ are the Pauli matrices. 

The generators of $\bG$ are normalized as 
\be
 \tr(T^IT^J) = \frac{1}{2}\dlt^{IJ}, 
\ee
and the structure constants are defined as 
\be
 \sbk{T^I,T^J} = iC_{IJK}T^K.  \label{def:C_IJK}
\ee

The spinorial representation of the $SO(5)$ generators is given by 
\be
 T^{\aL} \equiv \frac{1}{2}\mtrx{\sgm_{\aL}}{}{}{0_2}, \;\;\;\;\;
 T^{\aR} \equiv \frac{1}{2}\mtrx{0_2}{}{}{\sgm_{\aR}}, \;\;\;\;\;
 T^{\hat{a}} \equiv \frac{i}{2\sqrt{2}}\mtrx{}{\sgm_{\hat{a}}}
 {-\sgm_{\hat{a}}^\dagger}{}, 
 \label{SO_generator}
\ee
where $\sgm_{\hat{a}}\equiv(\vec{\sgm},-i1_2)$. 
Here $T^{\aL,\aR}$ ($\aL,\aR=1,2,3$) and $T^{\hat{a}}$ ($\hat{a}=1,2,3,4$) 
are the generators of $SO(4)\sim\suL\times\suR$ 
and $SO(5)/SO(4)$, respectively.

\section{Correction to the $WWHH$ coupling} \label{cal_WWHH}
Here we calculate the contribution of Fig.~\ref{ad_diagram} 
to the quartic coupling~$\lmd^{2\;\rm bare}_{WWHH}$ 
by the KK analysis. 
It is estimated as 
\be
 \dlt\lmd^2_{WWHH} \simeq -4\sum_{n=1}^\infty
 \frac{\lmd^2_{WW_nH}}{m_W^{(n)2}}, \label{exp_dltlmd}
\ee
where $m_W^{(n)}$ is a mass of the $n$-th KK excitation mode 
of the $W$ boson, and $\lmd_{WW_nH}$ is a trilinear coupling 
appearing in the 4D Lagrangian as 
\be
 \cL^{\rm int}_4 = \sum_n \lmd_{WW_nH}\brkt{W^{(0)\dagger}_\mu W^{(n)\mu}+
 W^{(n)\dagger}_\mu W^{(0)\mu}}H^{(0)}+\cdots.
\ee
Here $W_\mu^{(n)}$ and $H^{(0)}$ denote the $n$-th KK mode of the $W$ boson 
and the fluctuation zero-mode around the Higgs VEV, respectively. 
The numerical factor in (\ref{exp_dltlmd}) is a statistical factor 
of the Feynmann diagram in Fig.~\ref{ad_diagram} and 
the factor~$1/m^{(n)2}_W$ comes from the propagator of $W_\mu^{(n)}$. 
Following the procedure of Ref.~\cite{SH}, the coupling~$\lmd_{WW_nH}$ 
is expressed as the following overlap integral of the mode functions. 
\bea
 \lmd_{WW_nH} \eql \frac{g_A k}{2}\int_1^{\zp}\frac{dz}{z}
 \tl{h}^{\hat{4}}_{\vph,0}\left\{
 \tl{h}^{\hat{\pm}}_{A,0}\der_z\brkt{\tl{h}^{\chL}_{A,n}-\tl{h}^{\chR}_{A,n}}
 -\der_z\tl{h}^{\hat{\pm}}_{A,0}
 \brkt{\tl{h}^{\chL}_{A,n}-\tl{h}^{\chR}_{A,n}} \right. \nonumber\\
 && \hspace{30mm} \left. 
 +\tl{h}^{\hat{\pm}}_{A,n}\der_z\brkt{\tl{h}^{\chL}_{A,0}-\tl{h}^{\chR}_{A,0}}
 -\der_z\tl{h}^{\hat{\pm}}_{A,n}\brkt{\tl{h}^{\chL}_{A,0}-\tl{h}^{\chR}_{A,0}}
 \right\}, \label{lmd:overlap}
\eea
where $\zp\equiv e^{k\pi R}$ is the warp factor, 
\be
 \tl{h}^{\hat{4}}_{\vph,0}(z) = \sqrt{\frac{2}{k(\zp^2-1)}}z
\ee
is the mode function of the Higgs field~$H^{(0)}$, and 
\bea
 \tl{h}^{\chL}_{A,n}(z) \eql C^{\chL}_{A,n}zF_{1,0}(\lmd_n z,\lmd_n\zp), 
 \nonumber\\
 \tl{h}^{\chR}_{A,n}(z) \eql C^{\chR}_{A,n}zF_{1,0}(\lmd_n z,\lmd_n\zp), 
 \nonumber\\
 \tl{h}^{\hat{\pm}}_{A,n}(z) \eql C^{\hat{\pm}}_{A,n}
 zF_{1,1}(\lmd_n z,\lmd_n\zp) 
\eea
are those of $W^{(n)}_\mu$. 
Here the functions~$F_{\alp,\bt}(u,v)$ are defined in terms of 
the Bessel functions as 
\be
 F_{\alp,\bt}(u,v) \equiv J_\alp(u)Y_\bt(v)-Y_\alp(u)J_\bt(v), 
\ee
and $\lmd_n\equiv m_W^{(n)}/k$ is the mass eigenvalues determined by 
\be
 F_{1,0}\brc{\pi^2\lmd_n^2\zp F_{0,0}F_{1,1}-2\sin^2\bthH} = 0. 
\ee
Here and henceforth, $F_{\alp,\bt}$ without the argument denotes 
$F_{\alp,\bt}(\lmd_n,\lmd_n\zp)$. 
The coefficients~$(C^{\chL}_{A,n},C^{\chR}_{A,n},C^{\hat{\pm}}_{A,n})$ 
are given as follows. 

\noindent
{\bf \underline{Case 1}:}  The eigenvalue~$\lmd_n$ is determined 
 by $F_{1,0} = 0$.
\bea
 C^{\chL}_{A,n} \eql (1-\cos\bthH)\hat{C}^{(1)}_n, \nonumber\\
 C^{\chR}_{A,n} \eql -(1+\cos\bthH)\hat{C}^{(1)}_n, \nonumber\\
 C^{\hat{\pm}}_{A,n} \eql 0, 
\eea
where 
\be
 \hat{C}^{(1)}_n \equiv \frac{\sqrt{k}}{\sqrt{1+\cos^2\bthH}}
 \brc{\frac{4}{\pi^2\lmd_n^2}-F_{0,0}^2}^{-1/2}. 
\ee

\noindent
{\bf \underline{Case 2}:}  The eigenvalue~$\lmd_n$ is determined 
 by $\pi^2\lmd_n^2\zp F_{0,0}F_{1,1} = 2\sin^2\bthH$.
\bea
 C^{\chL}_{A,n} \eql (1+\cos\bthH)\hat{C}^{(2)}_n, \nonumber\\
 C^{\chR}_{A,n} \eql (1-\cos\bthH)\hat{C}^{(2)}_n, \nonumber\\
 C^{\hat{\pm}}_{A,n} \eql -\sqrt{2}\sin\bthH\frac{F_{1,0}}{F_{1,1}}
 \hat{C}^{(2)}_n, 
\eea
where 
\be
 \hat{C}^{(2)}_n \equiv \frac{\sqrt{k}}{\sqrt{1+\cos^2\bthH}}
 \brc{\frac{4}{\pi^2\lmd_n^2}+\frac{\pi^2\lmd_n^2\zp^2F_{1,0}^2F_{0,0}^2}
 {\sin^2\bthH(1+\cos^2\bthH)}-\frac{2F_{1,0}^2}{1+\cos^2\bthH}
 -\frac{2F_{0,0}^2}{\sin^2\bthH}}^{-1/2}. 
\ee
The detailed derivation of the above expressions are provided 
in Ref.~\cite{HS1}. 

When the warp factor~$\zp$ is large enough, 
the mode function of the lowest mode (\ie, the $W$ boson mode) is 
approximated as 
\be
 \tl{h}^{\chL}_{A,0}(z) \simeq \frac{1+\cos\bthH}{2\sqrt{\pi R}}, 
 \;\;\;\;\;
 \tl{h}^{\chR}_{A,0}(z) \simeq \frac{1-\cos\bthH}{2\sqrt{\pi R}}, 
 \;\;\;\;\;
 \tl{h}^{\hat{\pm}}_{A,0}(z) \simeq -\frac{\sin\bthH}{\sqrt{2\pi R}}
 \brkt{1-\frac{z^2}{\zp^2}}. 
\ee
Then the overlap integral~(\ref{lmd:overlap}) is simplified as 
\bea
 \lmd^{(1)}_{WW_nH} \sqa -\frac{4g_A\sqrt{k}\hat{C}^{(1)}_n\sin\bthH}
 {\lmd_n\zp^3\sqrt{\pi R}}\brkt{\frac{4}{\pi\lmd_n^2}+F_{0,0}}, 
 \nonumber\\
 \lmd^{(2)}_{WW_nH} \sqa -\frac{g_A\sqrt{k}\hat{C}^{(2)}_n\sin\bthH
 \cos\bthH}{\zp^3\sqrt{\pi R}}\brc{\brkt{1-\frac{8}{\lmd_n^2}}F_{1,0}
 +\frac{16}{\pi\lmd_n^3}+\frac{4F_{0,0}}{\lmd_n}},
\eea
where the superscript~$(1)$ or $(2)$ denotes 
that $W_\mu^{(n)}$ belongs to Case~1 or Case~2, respectively. 
We have used the formulae collected in the appendix~C in Ref.~\cite{HS1}. 

The dominant contribution in (\ref{exp_dltlmd}) 
comes from the modes which satisfy $m_W^{(n)}\ll k$ 
(or $\lmd_n\ll 1$). 
Thus we focus on such modes in the following. 
Then $F_{\alp,\bt}$ are approximated as 
\be
 F_{0,0} \simeq -\frac{2\ln\lmd_n}{\pi}J_0(\lmd_n\zp), 
 \;\;\;\;\;
 F_{1,0} \simeq \frac{2}{\pi\lmd_n}J_0(\lmd_n\zp), \;\;\;\;\;
 F_{1,1} \simeq \frac{2}{\pi\lmd_n}J_1(\lmd_n\zp). 
 \label{ap_Fs}
\ee
Making use of these approximations, we can obtain a simple expression 
for $\dlt\lmd_{WWHH}^2$. 

In Case~1, the mass eigenvalues~$\lmd_n$ satisfy 
$J_0(\lmd_n\zp)\simeq 0$ and 
\bea
 \dlt\lmd_{WWHH}^{(1)2} \sqa -4\sum_n\frac{\lmd^{(1)2}_{WW_nH}}{m_W^{(n)2}}
 \nonumber\\
 \sqa -4\sum_n\frac{16g_A^2\sin^2\bthH}{(1+\cos^2\bthH)\lmd_n^4\zp^6\pi R}
 \brkt{\frac{4}{\pi\lmd_n^2}+F_{0,0}}^2
 \brkt{\frac{4}{\pi^2\lmd_n^2}-F_{0,0}^2}^{-1} \nonumber\\
 \sqa -4\frac{\bar{g}^2\sin^2\bthH}{1+\cos^2\bthH}
 \sum_n\brkt{\frac{2}{\lmd_n\zp}}^6, 
\eea
where $\bar{g}\equiv g_A/\sqrt{\pi R}$. 

In Case~2, the equation that determines $\lmd_n$ 
is approximated as 
\be
 J_0(\lmd_n\zp)J_1(\lmd_n\zp) \simeq 
 \frac{\sin^2\bthH}{2\lmd_n\zp\ln\lmd_n} \ll 1. 
\ee
Thus $\lmd_n$ satisfy 
\bea
 J_0(\lmd_n\zp) \ll J_1(\lmd_n\zp) \simlt \cO(1), \;\;\;\;\;
 (n=2m+1) \nonumber\\
 J_1(\lmd_n\zp) \ll J_0(\lmd_n\zp) \simlt \cO(1), \;\;\;\;\;
 (n=2m+2) \label{rel_Js}
\eea
for $m=0,1,2,\cdots$. 
In this case, 
\bea
 \dlt\lmd_{WWHH}^{(2)2} \sqa -4\sum_n\frac{\lmd^{(2)2}_{WW_nH}}
 {m_W^{(n)2}} \nonumber\\
 \sqa -4\sum_n\frac{g_A^2\sin^2\bthH\cos^2\bthH}
 {(1+\cos^2\bthH)\lmd_n^2\zp^6\pi R}\brc{\brkt{1-\frac{8}{\lmd_n^2}}F_{1,0}
 +\frac{16}{\pi\lmd_n^2}+\frac{4F_{0,0}}{\lmd_n}}^2 \nonumber\\
 && \hspace{10mm} \cdot
 \brc{\frac{4}{\pi^2\lmd_n^2}+\frac{\pi^2\lmd_n^2\zp^2F_{1,0}^2F_{0,0}^2}
 {\sin^2\bthH(1+\cos^2\bthH)}-\frac{2F_{1,0}^2}{1+\cos^2\bthH}
 -\frac{2F_{0,0}^2}{\sin^2\bthH}}^{-1/2} \nonumber\\
 \sqa -4\frac{\bar{g}^2\sin^2\bthH\cos^2\bthH}{1+\cos^2\bthH}
 \sum_n\brkt{\frac{2}{\lmd_n\zp}}^6\brc{1-J_0(\lmd_n\zp)}^2 \nonumber\\
 && \hspace{10mm} \cdot
 \brc{1-\frac{2J_0^2(\lmd_n\zp)}{1+\cos^2\bthH}
 +\frac{J_0^2(\lmd_n\zp)\sin^2\bthH}{(1+\cos^2\bthH)J_1^2(\lmd_n)}}^{-1} 
 \nonumber\\
 \sqa -4\frac{\bar{g}^2\sin^2\bthH\cos^2\bthH}{1+\cos^2\bthH}
 \sum_m\brkt{\frac{2}{\lmd_{2m+1}\zp}}^6. 
\eea
We have used (\ref{rel_Js}) at the last step. 

Therefore (\ref{exp_dltlmd}) is estimated as 
\bea
 \dlt\lmd_{WWHH}^2 \eql \dlt\lmd_{WWHH}^{(1)2}+\dlt\lmd_{WWHH}^{(2)2}
 \nonumber\\
 \sqa -4\frac{\bar{g}^2\sin^2\bthH}{1+\cos^2\bthH}
 \sum_n\brkt{\frac{2}{x_n}}^6
 -4\frac{\bar{g}^2\sin^2\bthH\cos^2\bthH}{1+\cos^2\bthH}
 \sum_n\brkt{\frac{2}{x_n}}^6 \nonumber\\
 \eql -4\bar{g}^2\sin^2\bthH\sum_n\brkt{\frac{2}{x_n}}^6 \nonumber\\
 \eql -\frac{4}{3}\bar{g}^2\sin^2\bthH, 
\eea
where $x_n$ are zeros of $J_0(x)$. 
In the last equality, we have used the formula
\be
 \sum_n\brkt{\frac{2}{x_n}}^6 = \frac{1}{3}. 
\ee


\end{document}